\pdfoutput=1

\documentclass[12pt]{article}
\usepackage{acronym}

\usepackage{a4wide}
\usepackage{scalefnt}
\usepackage{bold-extra}
\usepackage{textcomp}
\usepackage{amsmath}
\usepackage{amssymb}
\usepackage{upgreek}
\usepackage{dsfont}
\usepackage[labelformat=simple]{subcaption}
\usepackage{graphicx}
\usepackage{xspace}
\usepackage{url}
\usepackage{hyperref}
\usepackage{fancyhdr}
\usepackage{booktabs}
\usepackage{longtable}
\usepackage{nicefrac}
\usepackage[numbers,sort&compress]{natbib}
\usepackage{filemod}
\usepackage[affil-it]{authblk} \usepackage[usenames,dvipsnames]{color}
\usepackage[final]{showlabels}
\usepackage[utf8x]{inputenc}
\showlabels{cite}

\usepackage[usenames,dvipsnames]{color}
\fancypagestyle{firstpage}
{
  
  \fancyhead[R]{\myheaderline}
}
\newcounter{notecount}
\newcommand{\mbottom}{m_\mathrm{b}}

\newcommand{\note}[2]{ {\stepcounter{notecount}\sf%
    \footnotesize\color{red}[\arabic{notecount}$|${\color{blue}#1}$|$#2]}}
\renewcommand{\note}[2]{}
\acrodef{QCD}{Quantum Chromo Dynamics}
\acused{QCD} 
\acrodef{IBP}{integration-by-parts}
\acrodef{BSM}{beyond-the-\ac{SM}}
\acrodef{LHC}{Large Hadron Collider}
\acrodef{LO}{leading order}
\acrodef{NLO}{next-to-leading order}
\acrodef{NNLO}{next-to-next-to-leading order}
\acrodef{PDF}{parton density function}
\acrodef{SM}{Standard Model}

\newcommand{\mquark}{m_\mathrm{q}}
\newcommand{\mhiggs}{M_\mathrm{H}}
\newcommand{\myheaderline}{%
  \small\sf
  August 2020\\
  FR-PHENO-2020-012\\
  MITP/20-045
}

\makeatletter
\DeclareRobustCommand*{\bfseries}{%
   \not@math@alphabet\bfseries\mathbf
   \fontseries\bfdefault\selectfont
   \boldmath
}
\makeatother

\numberwithin{equation}{section}

\newcommand{\order}[1]{\ensuremath{\mathcal{O}(#1)}}
\newcommand{\ggh}{\ensuremath{ggH}\xspace}

\newcommand{\MSbar}{\ensuremath{\overline{\text{MS}}}\xspace}

\renewcommand\Re{\operatorname{Re}}
\renewcommand\Im{\operatorname{Im}}
\newcommand\Li{\operatorname{Li}}
\renewcommand\d{{\mathrm{d}}}
\newcommand{\lbracket}{{[}}
\newcommand{\rbracket}{{]}}
\newcommand{\ltbrace}{{\symbol{123}}}
\newcommand{\rtbrace}{{\symbol{125}}}

\newcommand{\xIInt}[1]{{\mathds{I}_{#1}}}
\newcommand{\IInt}[2]{{\xIInt{#1}(#2)}}
\newcommand{\xpFq}[2]{{{}_{#1} F_{#2}}}
\newcommand{\pFq}[5]{{\xpFq{#1}{#2}\left(\begin{matrix} {#3} \\ {#4} \end{matrix} ; {#5}\right)}}
\DeclareMathSymbol{\shortminus}{\mathbin}{AMSa}{"39}

\title{The analytic leading color contribution to the Higgs-gluon form factor in QCD at NNLO}

\author[1]{Mario Prausa}
\author[2]{Johann Usovitsch}
\affil[1]{Physikalisches Institut,
Albert-Ludwigs-Universit\"at,\protect\\
D-79085 Freiburg, Germany}
\affil[2]{PRISMA Cluster of Excellence, Institut f\"ur Physik,\protect\\Johannes Gutenberg-Universit\"at Mainz,\protect\\D-55099 Mainz, Germany}

\date{}

\begin{document}
\maketitle
\thispagestyle{firstpage}
\begin{abstract}
    We present an analytic three-loop result for the leading color contribution to the Higgs-gluon form factor in QCD.
    The leading color contribution is given at next-to-next-to-leading order by the $N_c^2$-term in QCD with $N_c$ colors.
    The main focus of this article lies on the evaluation of the relevant Feynman integrals with a special emphasis on the elliptic sector.
\end{abstract}


\section{Introduction}

A precise knowledge of the gluon-gluon-Higgs (\ggh) form factor with arbitrary quark masses and virtualities is essential for many Higgs physics applications.
For example, it is a main ingredient in the total cross sections for single and double Higgs production at the \ac{LHC}.

Therefore, the evaluation of the \ggh form factor in \ac{QCD} has a long history: The initial groundwork was done by completing the exact two-loop result for arbitrary quark masses in Refs.~\cite{Spira:1995rr,Harlander:2005rq,Anastasiou:2006hc,Aglietti:2006tp}.
First computations at three loop level go back to large-mass expansions in the top quark mass~\cite{Harlander:2009bw,Pak:2009bx}.
More recent results exploit the partial knowledge of the threshold behavior of the form factors~\cite{Grober:2017uho} and apply the techniques of Pad\'e approximations~\cite{Davies:2019nhm}, which in principle is sufficient for phenomenology.
Subsequently, the applicability of the Pad\'e approximations for top quark induced loops was proven shortly after in Ref.~\cite{Czakon:2020vql}.
Furthermore, using the result of Ref.~\cite{Czakon:2020vql}, the b-quark mass effects could be included exactly for the first time.
Aside from that, a large-mass expansion is already known at four loop level~\cite{Davies:2019wmk}.

In contrast to the purely numerical result of Ref.~\cite{Czakon:2020vql}, we choose a different approach in this article.
We want to continue our work started in Ref.~\cite{Harlander:2019ioe} towards a full analytic three-loop result for the \ggh form factor in QCD with a single massive quark by completing the calculation for the leading color contribution.
The quark mass dependence has been kept exactly in our calculation, in order to provide a result that is valid for all quark flavors.

Nowadays, the integration-by-parts reductions for three loop form factors are no longer a bottleneck and are manageable on a laptop within a few weeks.
The following evaluation of master integrals can be grouped into two categories.
The master integrals that can be expressed in terms of multiple polylogarithms are by now very well understood and their evaluation can be automated to a high degree.

Unfortunately, the computation of master integrals in the elliptic sector is still a big challenge.
Pioneering work in this area was done in Refs.~\cite{Broadhurst:1993mw,Berends:1993ee,Bauberger:1994nk,Bauberger:1994by,Bauberger:1994hx,Caffo:1998du,Laporta:2004rb,Kniehl:2005bc,Groote:2005ay,Groote:2012pa,Bailey:2008ib,MullerStach:2011ru,Adams:2013nia,Bloch:2013tra,Adams:2014vja,Adams:2015gva,Adams:2015ydq,Remiddi:2013joa,Bloch:2016izu,Groote:2018rpb}, where the simplest elliptic Feynman integral, the two-loop equal-mass sunrise diagram, was evaluated.
These initial works have incited an active research field into elliptic Feynman integrals~\cite{Adams:2017ejb,Bloch:2014qca,Remiddi:2016gno,Adams:2016xah,Bogner:2017vim,Adams:2018yfj,Sogaard:2014jla,Bonciani:2016qxi,vonManteuffel:2017hms,Primo:2017ipr,Ablinger:2017bjx,Bourjaily:2017bsb,Hidding:2017jkk,Passarino:2016zcd,Remiddi:2017har,Lee:2017qql,Broedel:2017kkb,Broedel:2017siw,Broedel:2018iwv,Lee:2018ojn,Adams:2018kez,Adams:2018bsn,Lee:2018jsw,Broedel:2018qkq,Broedel:2019hyg,Bogner:2019lfa,Broedel:2019kmn,Frellesvig:2019byn,Abreu:2019fgk}.

The master integrals of the leading color contribution to the \ggh form factor contain only a single elliptic sector.
In order to give analytic results for this sector a new class of iterated integrals has to be introduced with integration kernels depending on complete elliptic integrals of the first kind.
The results in this article can be evaluated numerically via an included implementation of these iterated integrals.

In section~\ref{sect:setup}, the computational setup and toolchain of the calculation is described.
Section~\ref{sect:masters} is devoted to the evaluation of the various types of master integrals.
In section~\ref{sect:result}, we present the final result of this paper: the leading color contribution at three-loop including explicitly the analytic dependence of the quark mass.
We are able to express the result in terms of three different classes of iterated integrals.
Finally, our conclusion and an outlook to possible future work is given in section~\ref{sect:conclusion}.

\section{Computational Setup} \label{sect:setup}
\begin{figure}
  \centering
  \begin{subfigure}[b]{.18\textwidth}
    \centering
    \includegraphics[scale=.23]{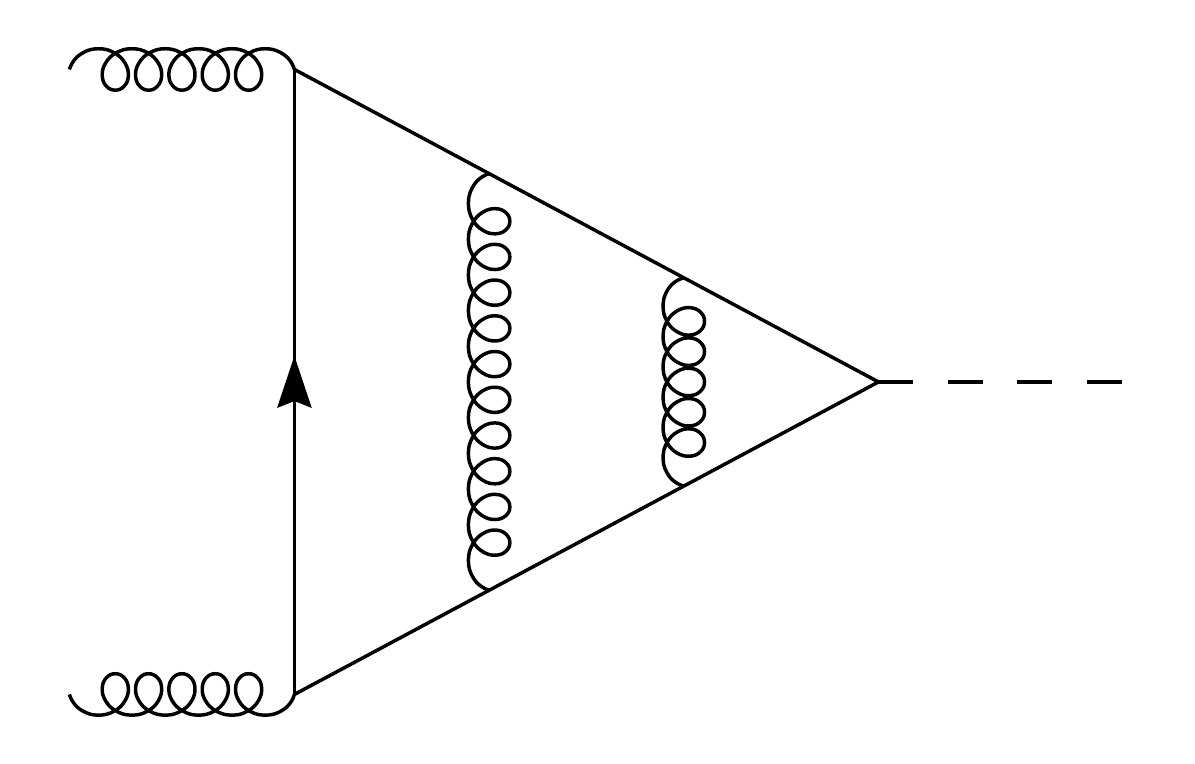}
    \caption{}
  \end{subfigure}%
  \begin{subfigure}[b]{.18\textwidth}
    \centering
    \includegraphics[scale=.23]{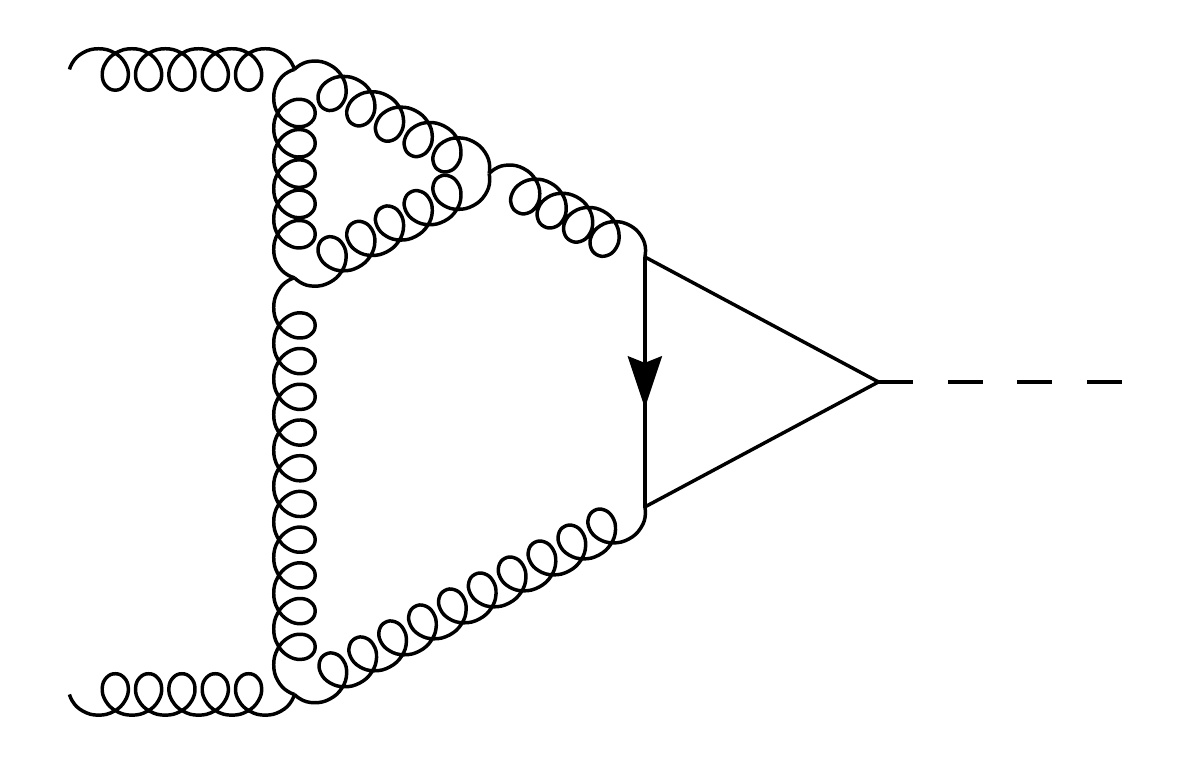}
    \caption{}
  \end{subfigure}%
  \begin{subfigure}[b]{.18\textwidth}
    \centering
    \includegraphics[scale=.23]{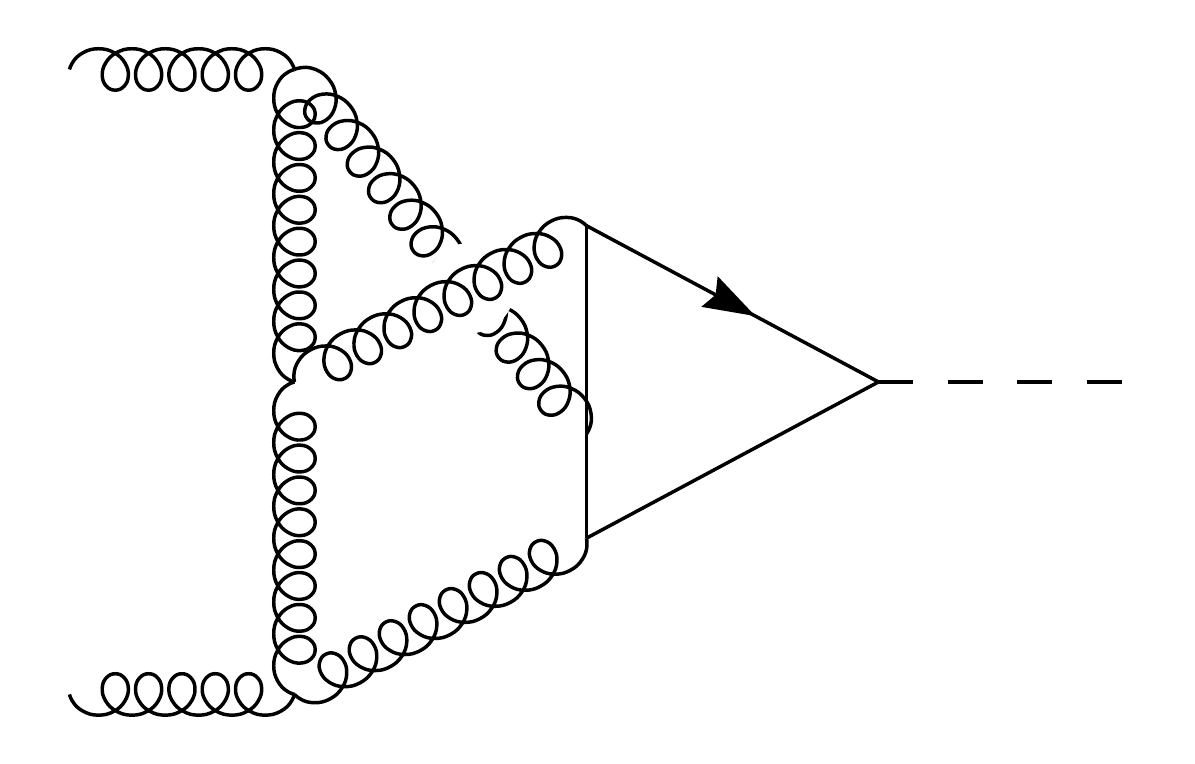}
    \caption{}
  \end{subfigure}%
  \begin{subfigure}[b]{.18\textwidth}
    \centering
    \includegraphics[scale=.23]{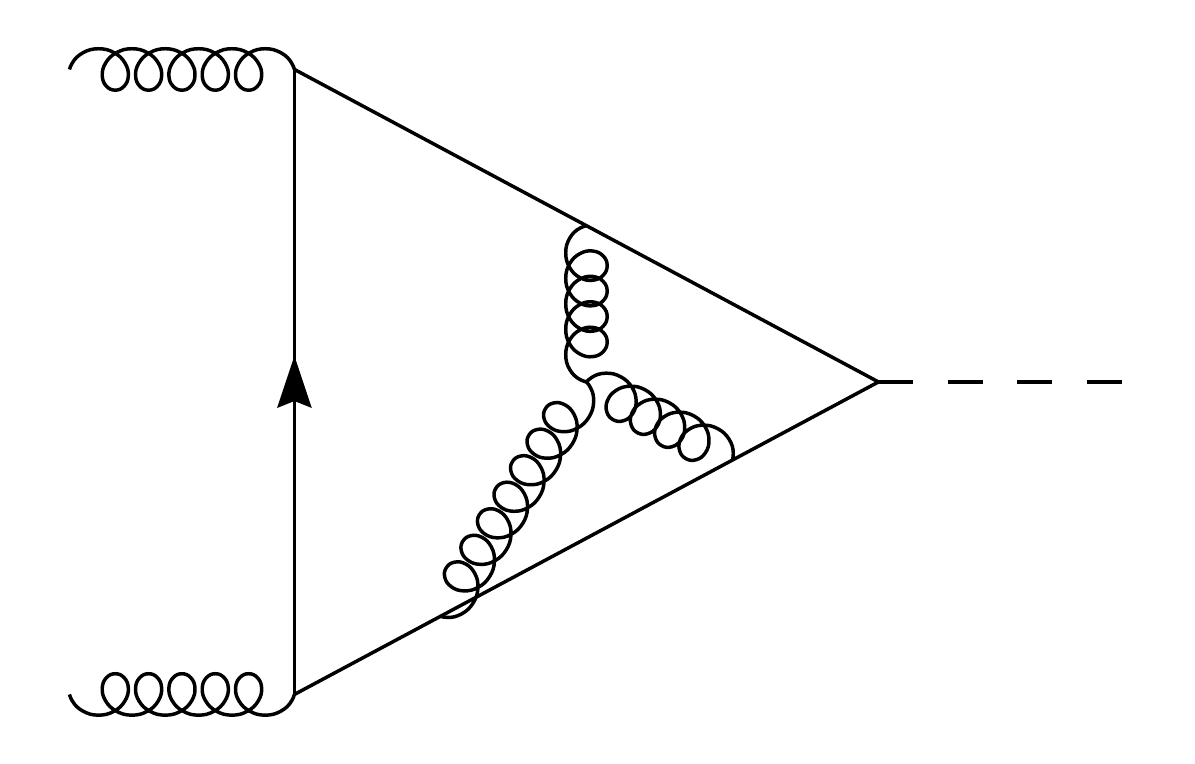}
    \caption{}
  \end{subfigure}%
  \begin{subfigure}[b]{.18\textwidth}
    \centering
    \includegraphics[scale=.23]{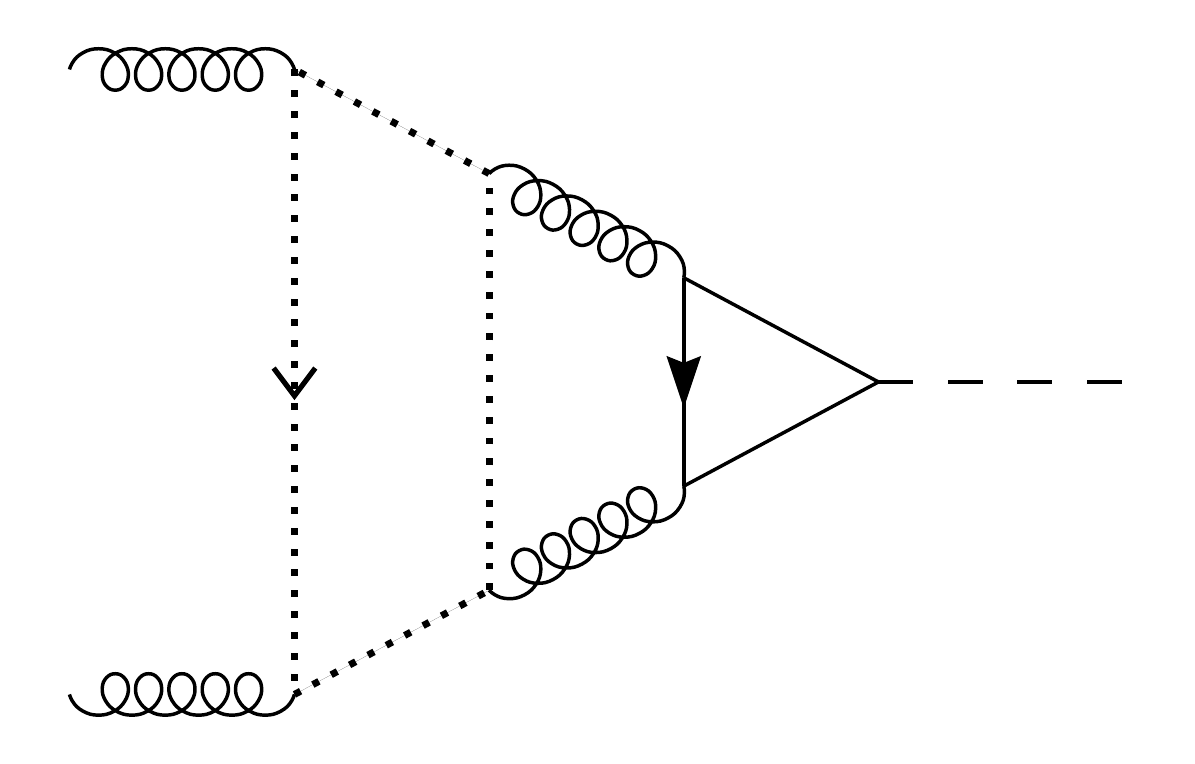}
    \caption{}
  \end{subfigure}%
  \\
  \begin{subfigure}[b]{.18\textwidth}
    \centering
    \includegraphics[scale=.23]{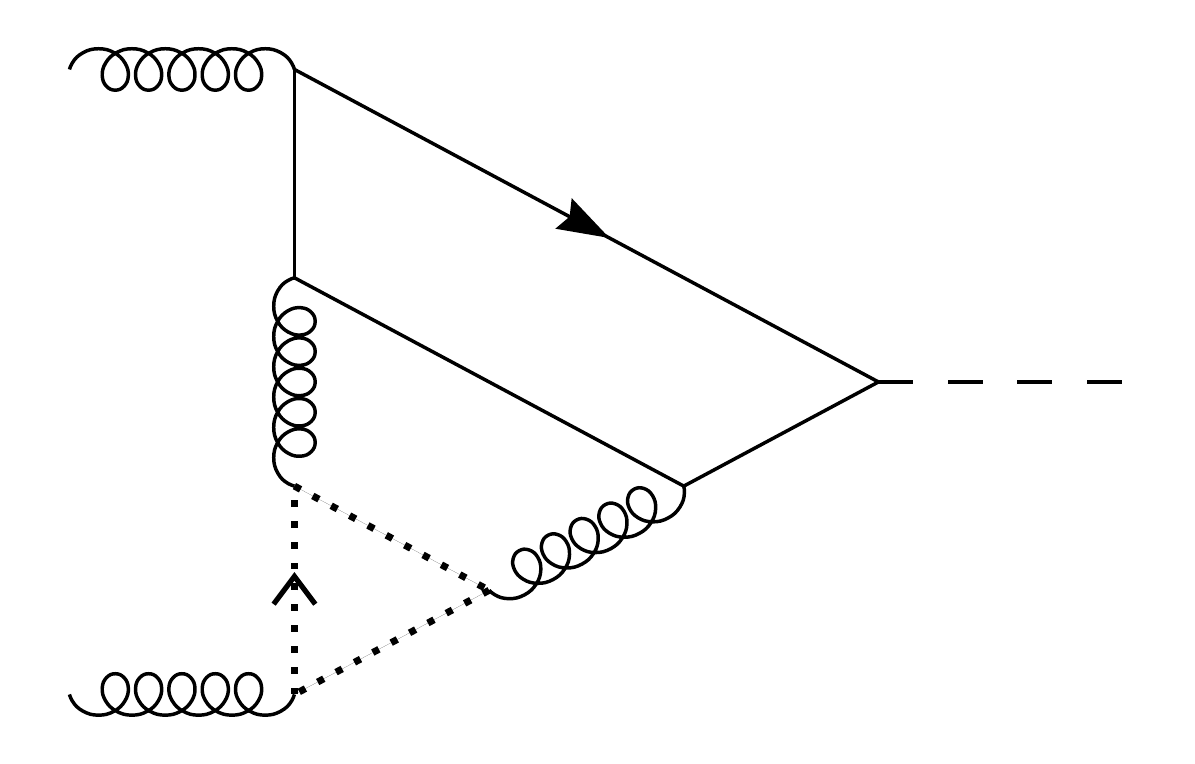}
    \caption{}
  \end{subfigure}%
  \begin{subfigure}[b]{.18\textwidth}
    \centering
    \includegraphics[scale=.23]{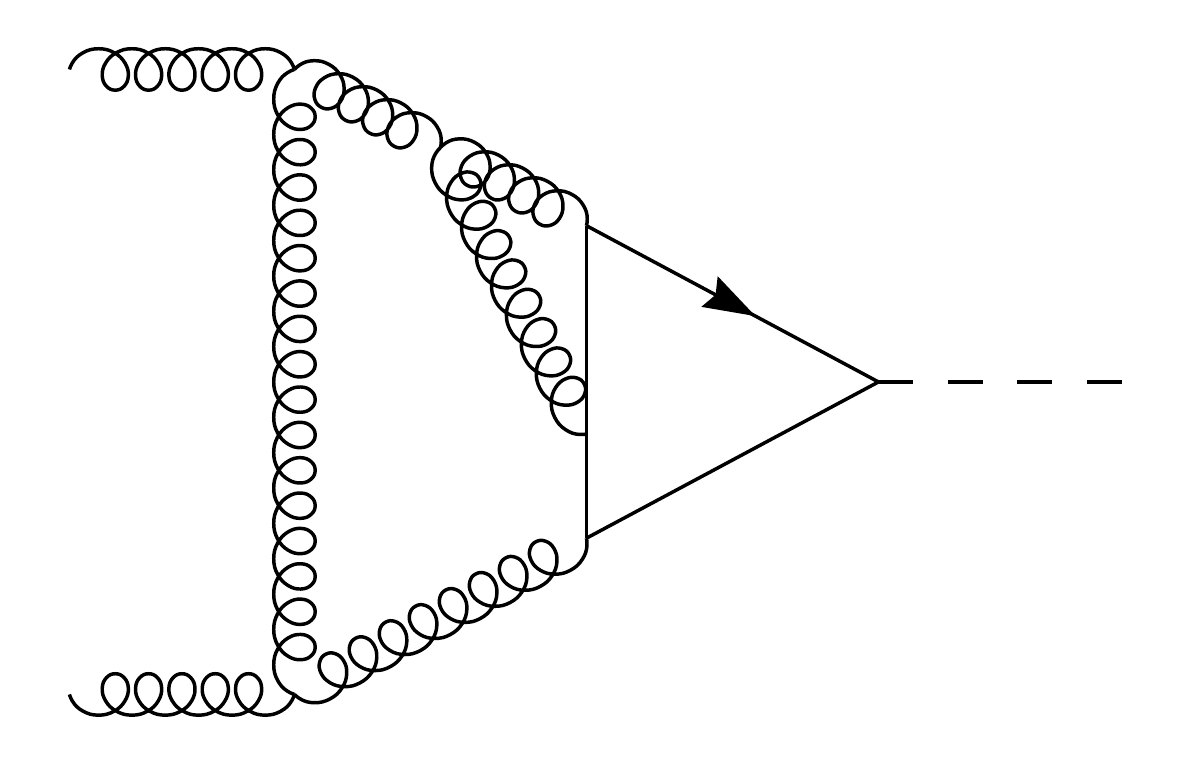}
    \caption{}
  \end{subfigure}%
  \begin{subfigure}[b]{.18\textwidth}
    \centering
    \includegraphics[scale=.23]{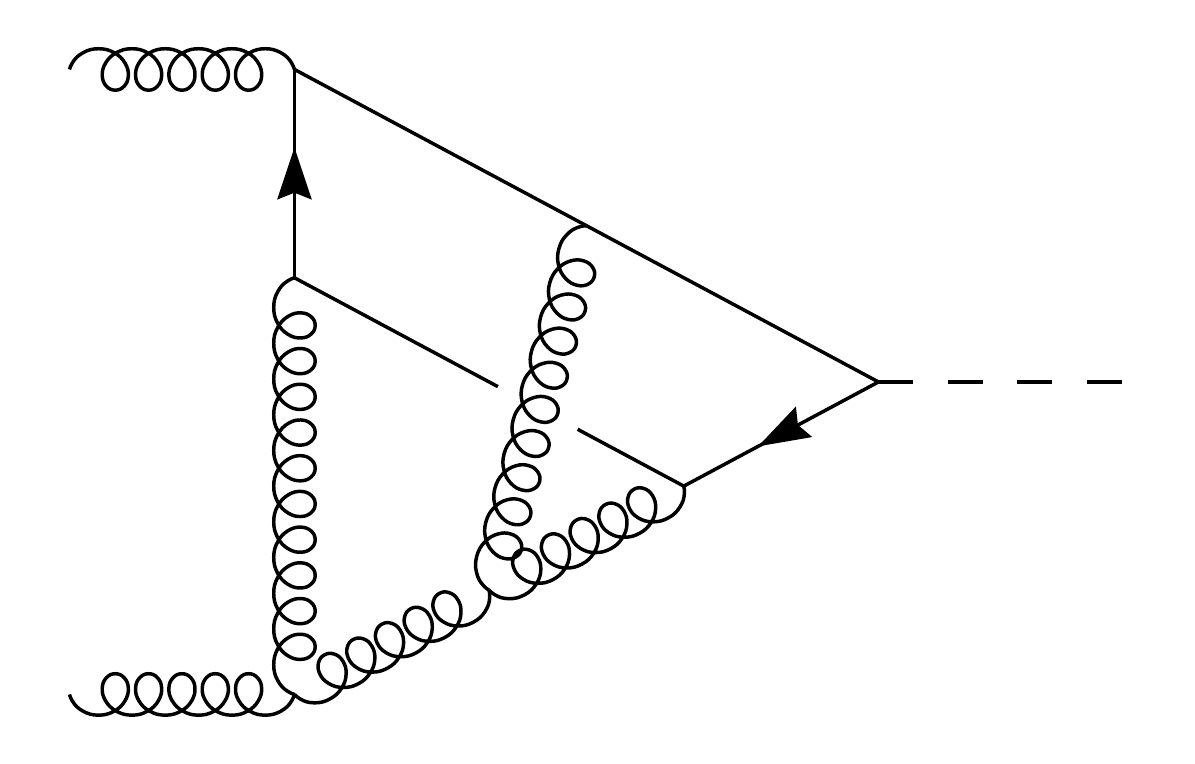}
    \caption{}
  \end{subfigure}%
  \begin{subfigure}[b]{.18\textwidth}
    \centering
    \includegraphics[scale=.23]{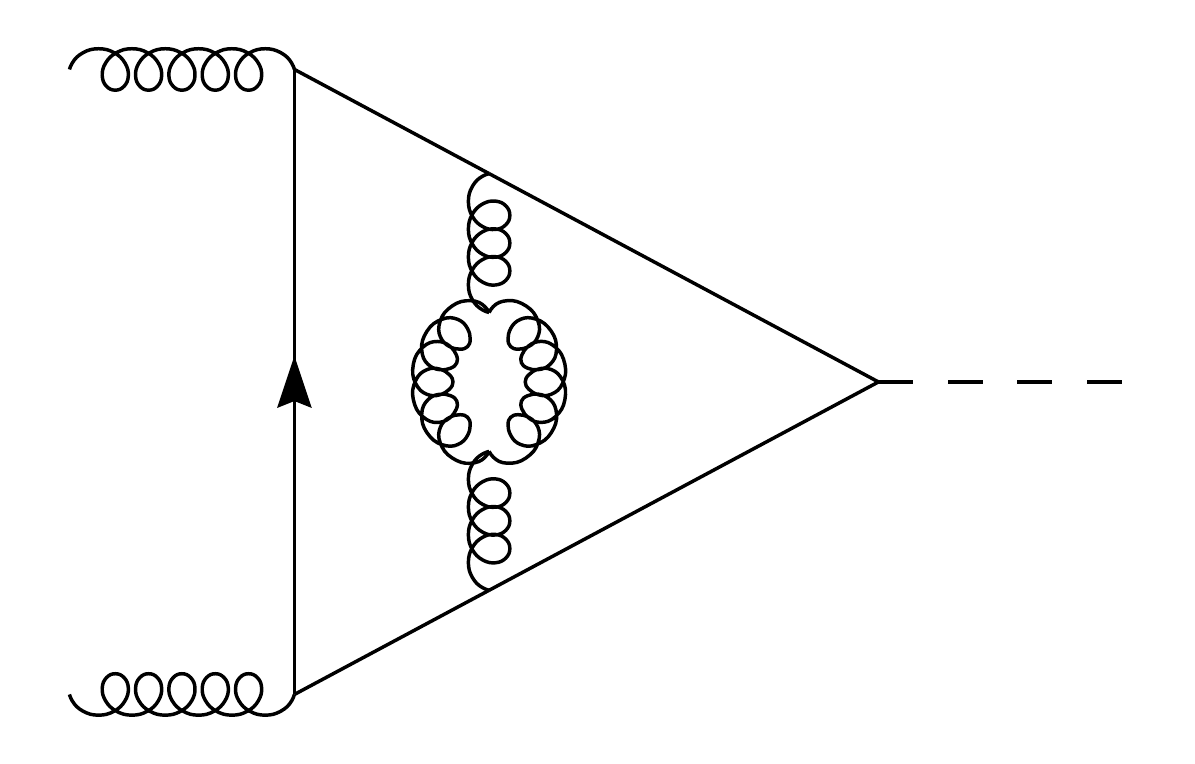}
    \caption{}
  \end{subfigure}%
  \begin{subfigure}[b]{.18\textwidth}
    \centering
    \includegraphics[scale=.23]{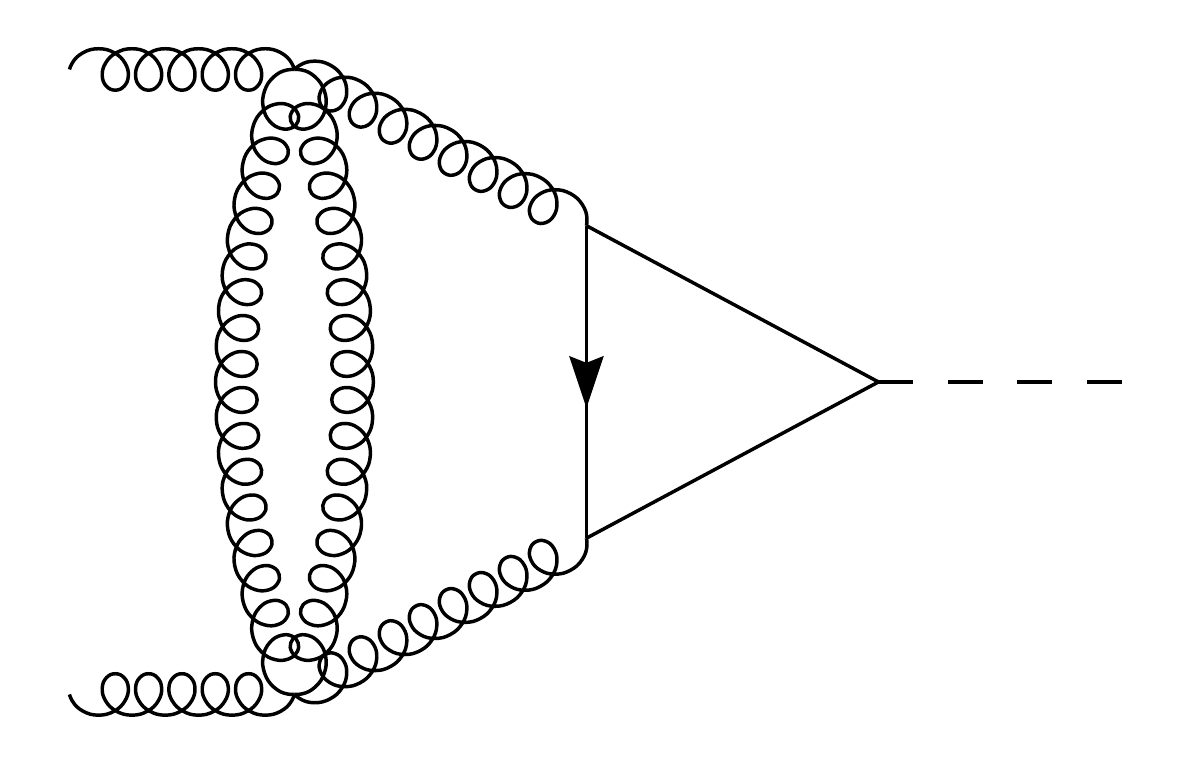}
    \caption{}
    \label{fig:feyndias:aah-first}
  \end{subfigure}%

  \caption{
    Sample Feynman diagrams to the leading color contribution to the \ggh form factor.
  }
  \label{fig:fdias}
\end{figure}%
The computational setup of the leading color contribution to the \ggh form factor consists of multiple steps.
At first, the complete set of Feynman diagrams in \ac{QCD} with two incoming gluons and one outgoing Higgs boson is generated with the computer program \texttt{qgraf}~\cite{Nogueira:1991ex}.
Sample diagrams are shown in fig.~\ref{fig:fdias}.
The output files of \texttt{qgraf} are further processed by \texttt{q2e}~\cite{Harlander:1997zb,Seidensticker:1999bb} in order to insert Feynman rules in Feynman gauge.
In a next step, the Feynman diagrams are mapped to a predefined set of topologies via the tool \texttt{exp}~\cite{Harlander:1997zb,Seidensticker:1999bb}, which generates output files for the computer algebra system \texttt{FORM}~\cite{Vermaseren:2000nd}.

The physical tensor structure of the Feynman amplitude ${\cal M}^{ab;\mu\nu}$ is given by
\begin{equation}
    {\cal M}^{ab;\mu\nu}
    =
    \delta^{ab} \left[q_2^\mu q_1^\nu - (q_1\cdot q_2) g^{\mu\nu}\right] C
    \,,
\end{equation}
where $q_1^{\mu}$ and $q_2^{\nu}$ are the components of the momenta of the external gluons.
The form factor $C$ can be projected out via
\begin{equation} \label{eq:projector}
    C
    =
    \frac{
        \delta^{ab}
        \left[
            (q_1\cdot q_2) g_{\mu\nu}
            -
            q_{2\mu} q_{1\nu}
        \right]
    }{
        N_A (q_1\cdot q_2)^2 (2-d)
    }
    {\cal M}^{ab;\mu\nu}
    \,,
\end{equation}
where $N_A$ is the number of gauge generators and $d=4-2\epsilon$ the number of space-time dimensions.

A custom \texttt{FORM}-code implements eq.~\eqref{eq:projector} and expresses $C$ in terms of scalar Feynman integrals.
The individual group theory factors of the Feynman diagrams are evaluated with the \texttt{FORM} package \texttt{color}~\cite{vanRitbergen:1998pn}, which expresses the color factors in terms of the Casimir invariants $C_F$, $C_A$ and $d_F^{abc} d_F^{abc}$.
In the QCD gauge group $SU(N_c)$ these Casimir invariants have the values
\begin{equation} \label{eq:sunc-color}
    C_F = \frac{T_F N_A}{N_c}
    \,, \quad
    C_A = 2T_F N_c
    \,, \quad
    d_F^{abc} d_F^{abc}
    =
    \frac{T_F^3 N_A}{2N_c} \left(N_c^2-4\right)
    \,,
\end{equation}
with $N_A = N_c^2-1$ and the normalization $T_F=1/2$.

The scalar Feynman integrals are then reduced to 409 master integrals via integration-by-parts relations~\cite{Tkachov:1981wb,Chetyrkin:1981qh} using Laporta's algorithm~\cite{Laporta:2001dd} implemented in the \texttt{C++} program \texttt{Kira}~\cite{Maierhoefer:2017hyi,Maierhofer:2018gpa}.

After inserting eq.~\eqref{eq:sunc-color}, the color structure in terms of $N_c$ is given by
\begin{equation} \label{eq:color-struct}
    C = \sum\limits_{k=-2}^2 N_c^k C^{(k)}
    \,.
\end{equation}
Since a complete analytical calculation of all $C^{(k)}$ in eq.~\eqref{eq:color-struct} is not feasible at the moment, we restrict ourselves to the leading color contribution $C^{(2)}$ in the following.
Only 196 master integrals contribute to this term.

\section{Master integrals} \label{sect:masters}
In order to study the kinematic dependency of master integrals the method of differential equations~\cite{Kotikov:1990kg,Kotikov:1991hm,Kotikov:1991pm,Gehrmann:1999as} turned out to be an indispensable tool.
The coupled system of differential equations can be simplified significantly if it is expressed in terms of an appropriate kinematic variable.
For our purposes, we choose in accordance with Ref.~\cite{Harlander:2019ioe} the variable $x$ related to the variable $z=\mhiggs^2/(4\mquark^2)+\mathrm{i}0$ by,
\begin{equation} \label{eq:variable-x}
  x = \frac{\sqrt{1-1/z}-1}{\sqrt{1-1/z}+1}\,,
  \quad
  z = -\frac{(1-x)^2}{4x}\,,
\end{equation}
with the mass $\mhiggs$ of the Higgs boson and the mass $\mquark$ of the massive quark\footnote{In Ref.~\cite{Harlander:2019ioe} the variable $z$ was named $\tau$.}.
Arranging all master integrals in a vector $\vec F(x,\epsilon)$, where the integrals are ordered by an increasing number of propagators in such a way that integrals of the same sector stay together, leads to a system of differential equations
\begin{equation}
    \frac{\d}{\d x} \vec F(x,\epsilon) = \hat M(x,\epsilon) \vec F(x,\epsilon)\,,
\end{equation}
where $\hat M(x,\epsilon)$ is a block-triangular matrix.
The differential equations of a certain sector can therefore be written as
\begin{equation} \label{eq:general-de}
    \frac{\d}{\d x} \vec f(x,\epsilon) = M(x,\epsilon) \vec f(x,\epsilon) + B(x,\epsilon) \vec g(x,\epsilon)\,,
\end{equation}
where $\vec f(x,\epsilon)$ contains only the master integrals of the sector under consideration and $\vec g(x,\epsilon)$ the master integrals with a smaller number of propagators.
Eq.~\eqref{eq:general-de} is the starting point for all the calculation methods described in the following subsections.
Solving the master integrals in $\vec F(x,\epsilon)$ sector by sector always allows us to consider $\vec g(x,\epsilon)$ in eq.~\eqref{eq:general-de} as already known.

The system of differential equations in eq.~\eqref{eq:general-de} fixes the master integrals up to some boundary conditions.
A suitable boundary condition for all master integrals under consideration is an expansion around $x\approx1$ corresponding to a large quark mass $\mquark$ compared to the Higgs mass $\mhiggs$.
It is straightforward to obtain such an asymptotic series using an expansion by subgraphs~\cite{Gorishnii:1989dd,Smirnov:1990rz,Smirnov:1994tg,Smirnov:2002pj} implemented in \texttt{exp}~\cite{Harlander:1997zb,Seidensticker:1999bb}.

The analytic results for the master integrals derived in the context of this article have all been checked against numerical results obtained by sector decomposition~\cite{Binoth:2000ps,Binoth:2003ak,Binoth:2004jv} at an euclidean point $x=0.01$ via the tool \texttt{FIESTA}~\cite{Smirnov:2015mct}.

The method treated in section~\ref{sect:canonical-x} is applicable to the vast majority of master integrals under consideration, i.e. the master integrals expressible in terms of generalized polylogarithms (GPLs) with argument $x$.
The two sectors described in section~\ref{sect:canonical-sqrtx} can also be expressed in terms of GPLs but require an argument of $\sqrt{x}$.
Section~\ref{sect:elliptic} covers the only elliptic sector present in the leading color contribution.
Section~\ref{sect:mixed} deals with two further sectors, where the homogenous parts of eq.~\eqref{eq:general-de} are of the type described in Section~\ref{sect:canonical-x} but the inhomogeneities $\vec g(x,\epsilon)$ contain integrals of the two other types.

\subsection{Canonical sectors in $x$} \label{sect:canonical-x}
\begin{figure}
  \centering
  \begin{subfigure}[b]{.18\textwidth}
    \centering
    \includegraphics[scale=.2]{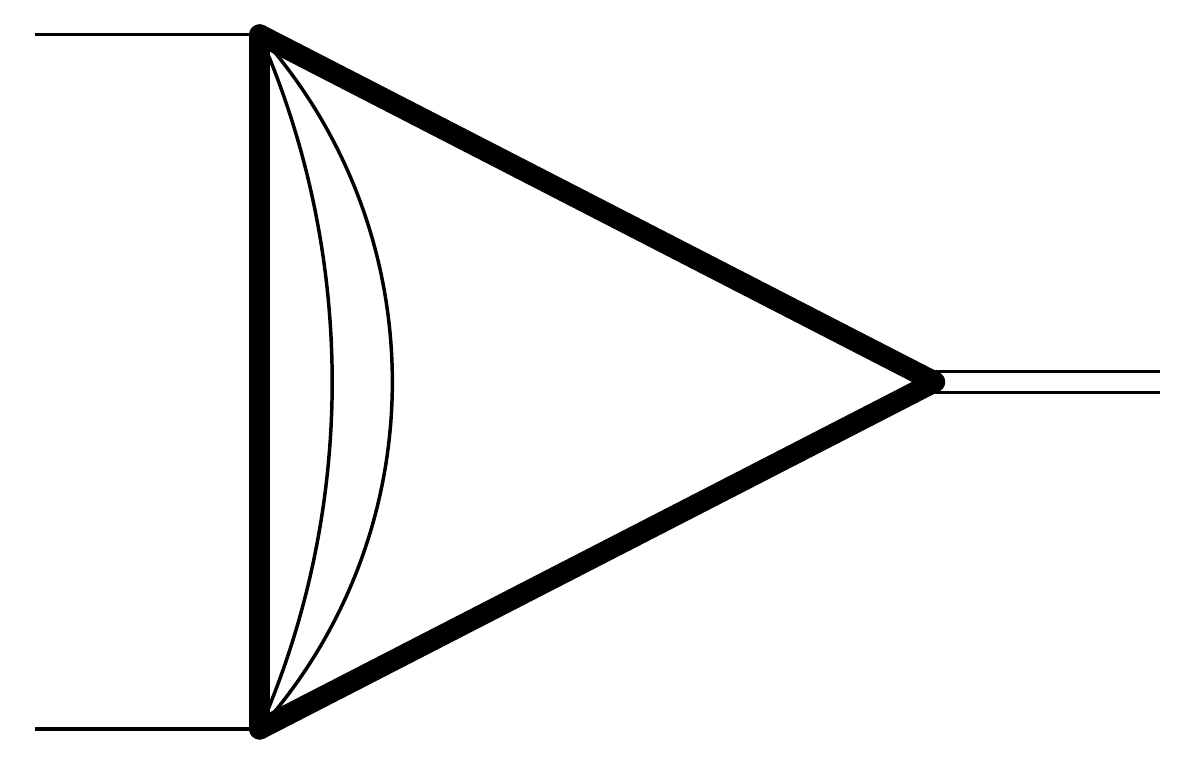}
    \caption{}
  \end{subfigure}%
  \begin{subfigure}[b]{.18\textwidth}
    \centering
    \includegraphics[scale=.2]{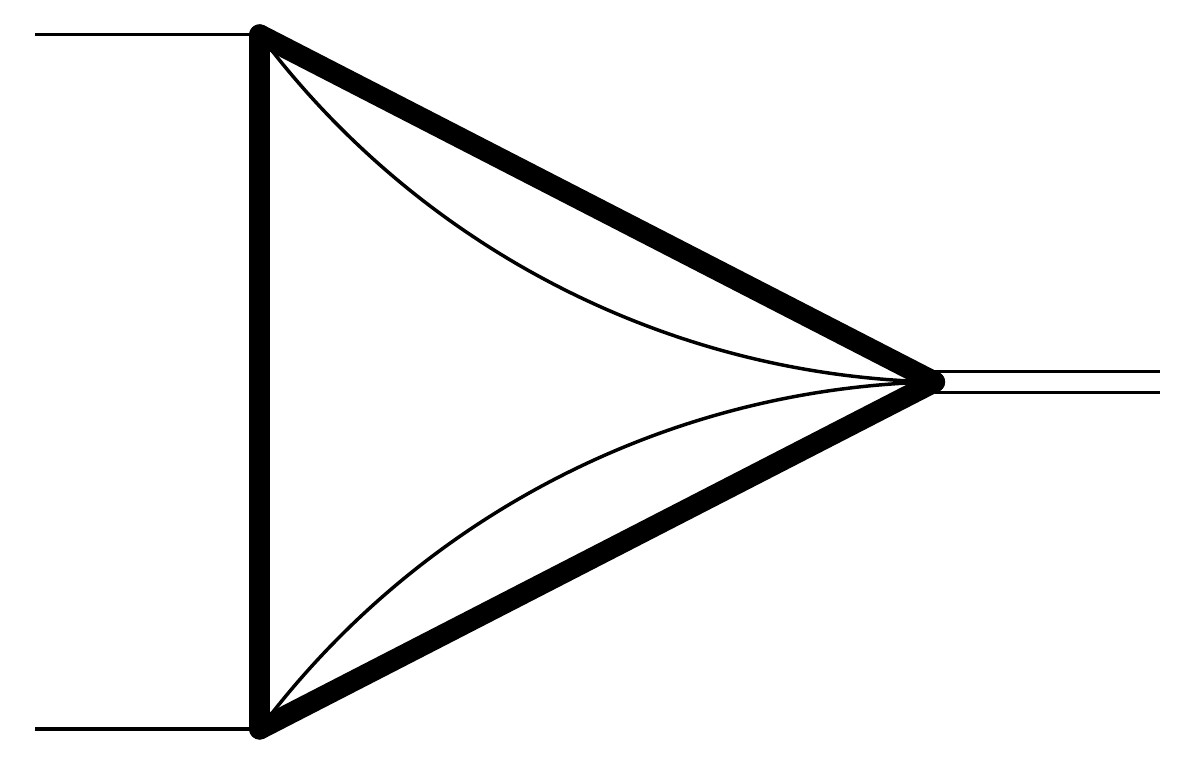}
    \caption{}
  \end{subfigure}%
  \begin{subfigure}[b]{.18\textwidth}
    \centering
    \includegraphics[scale=.2]{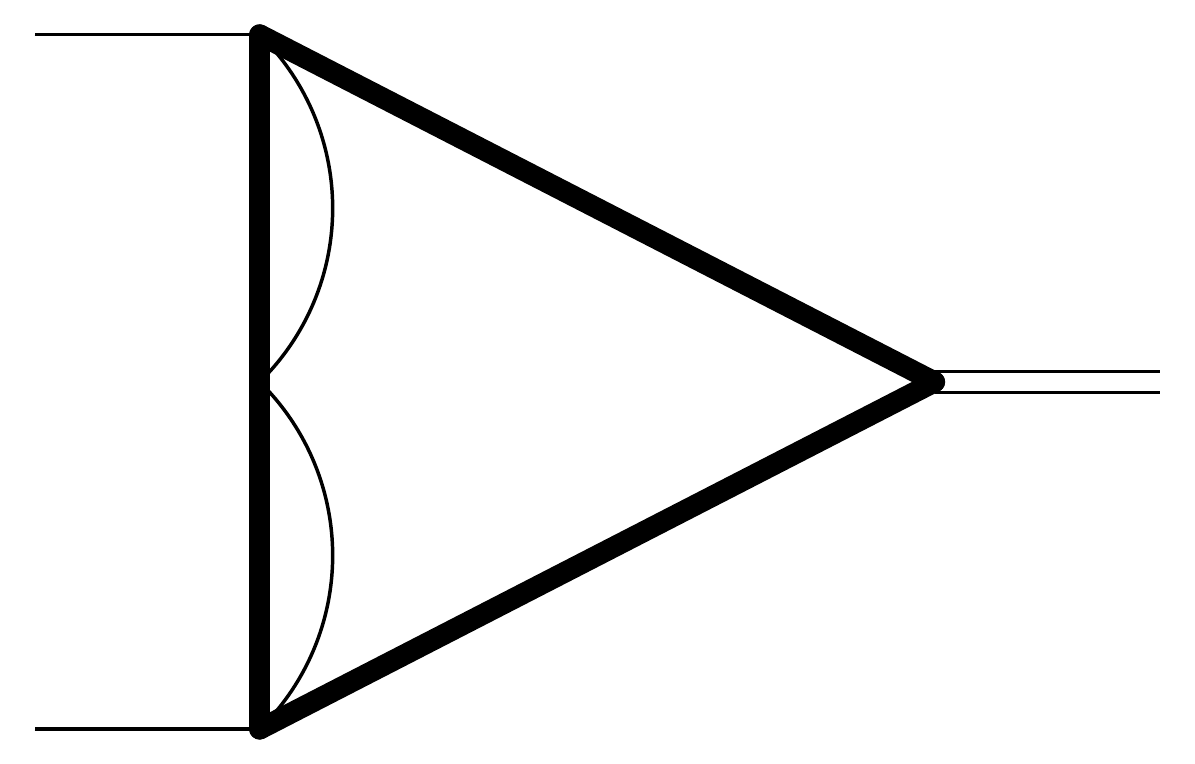}
    \caption{}
  \end{subfigure}%
  \begin{subfigure}[b]{.18\textwidth}
    \centering
    \includegraphics[scale=.2]{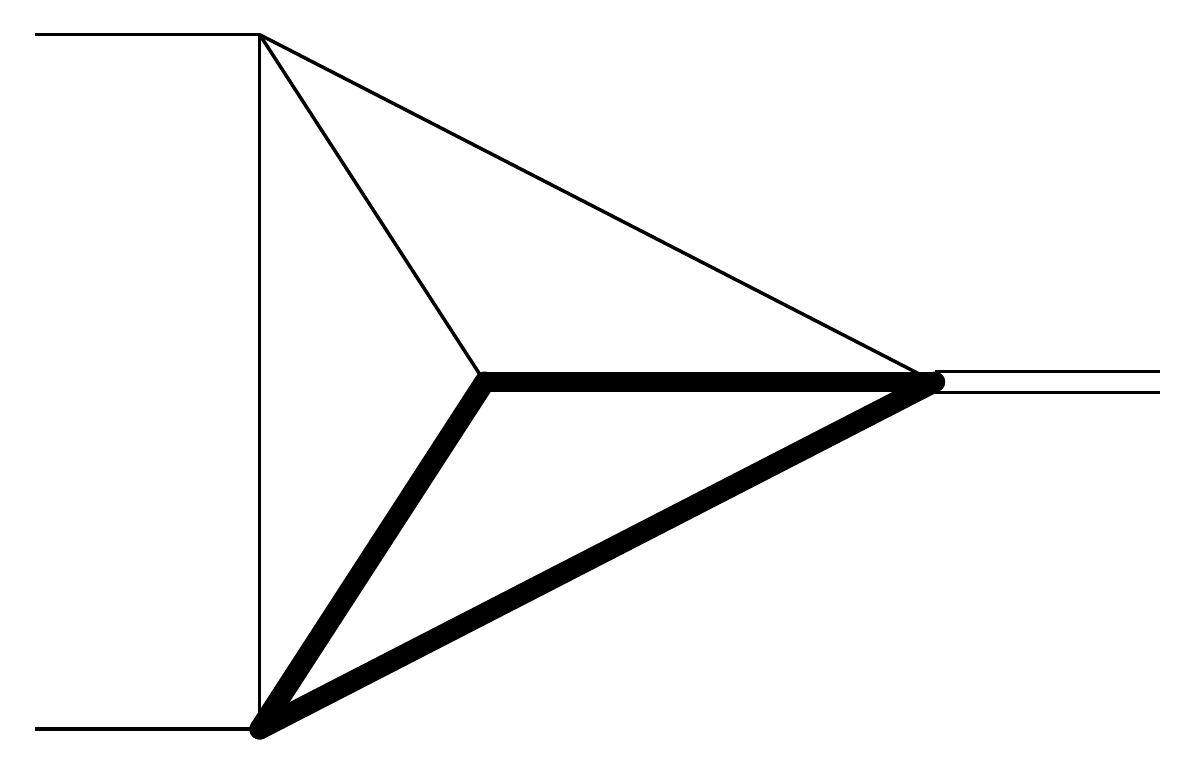}
    \caption{}
  \end{subfigure}%
  \begin{subfigure}[b]{.18\textwidth}
    \centering
    \includegraphics[scale=.2]{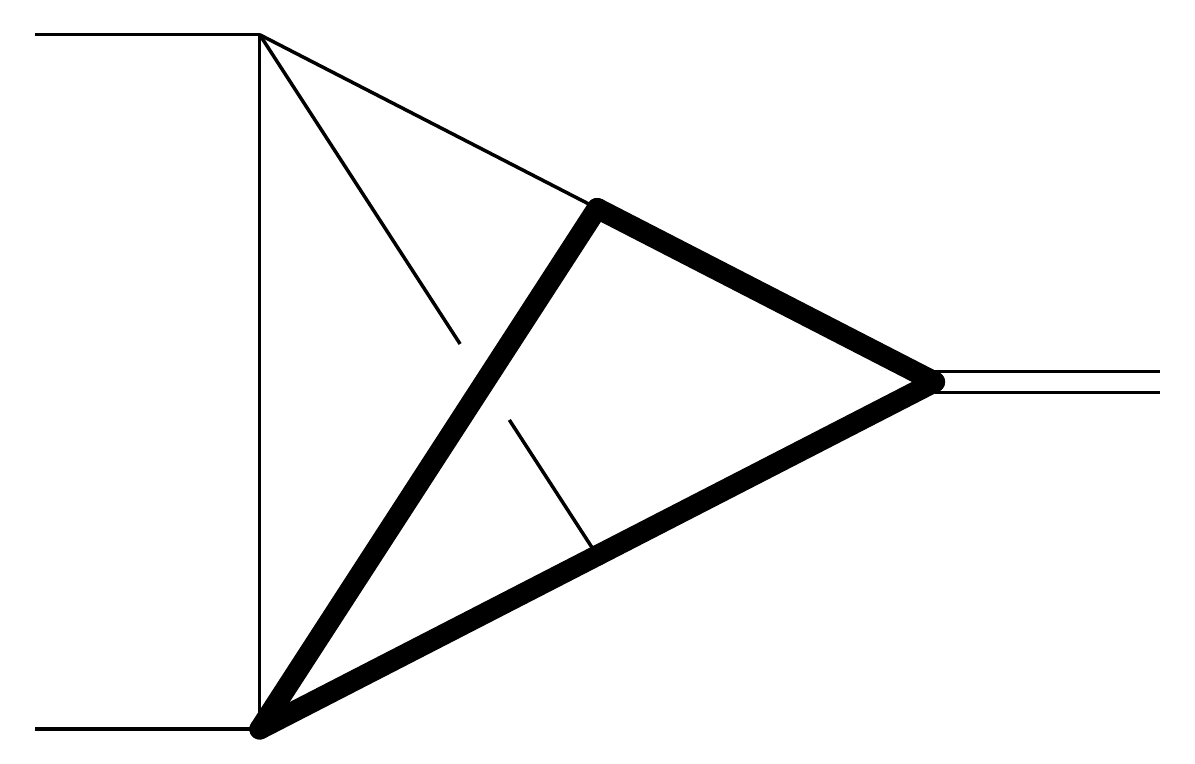}
    \caption{}
  \end{subfigure}%
  \\
  \begin{subfigure}[b]{.18\textwidth}
    \centering
    \includegraphics[scale=.2]{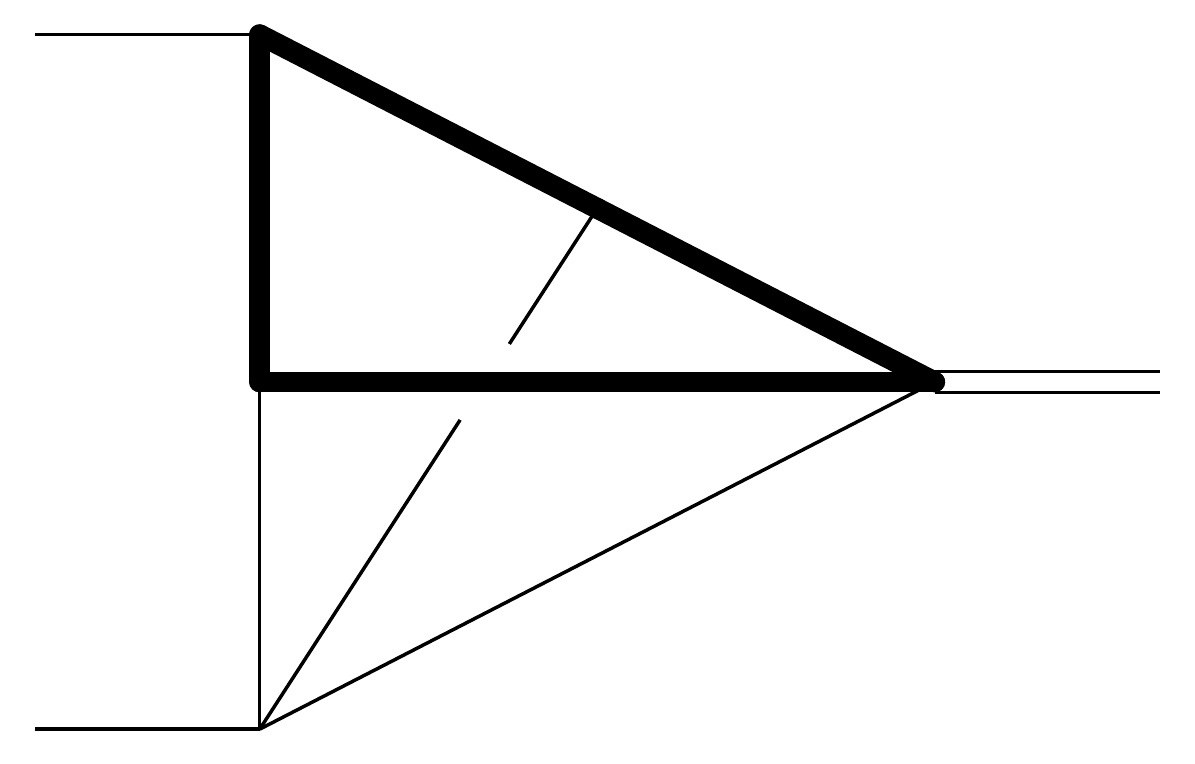}
    \caption{}
  \end{subfigure}%
  \begin{subfigure}[b]{.18\textwidth}
    \centering
    \includegraphics[scale=.2]{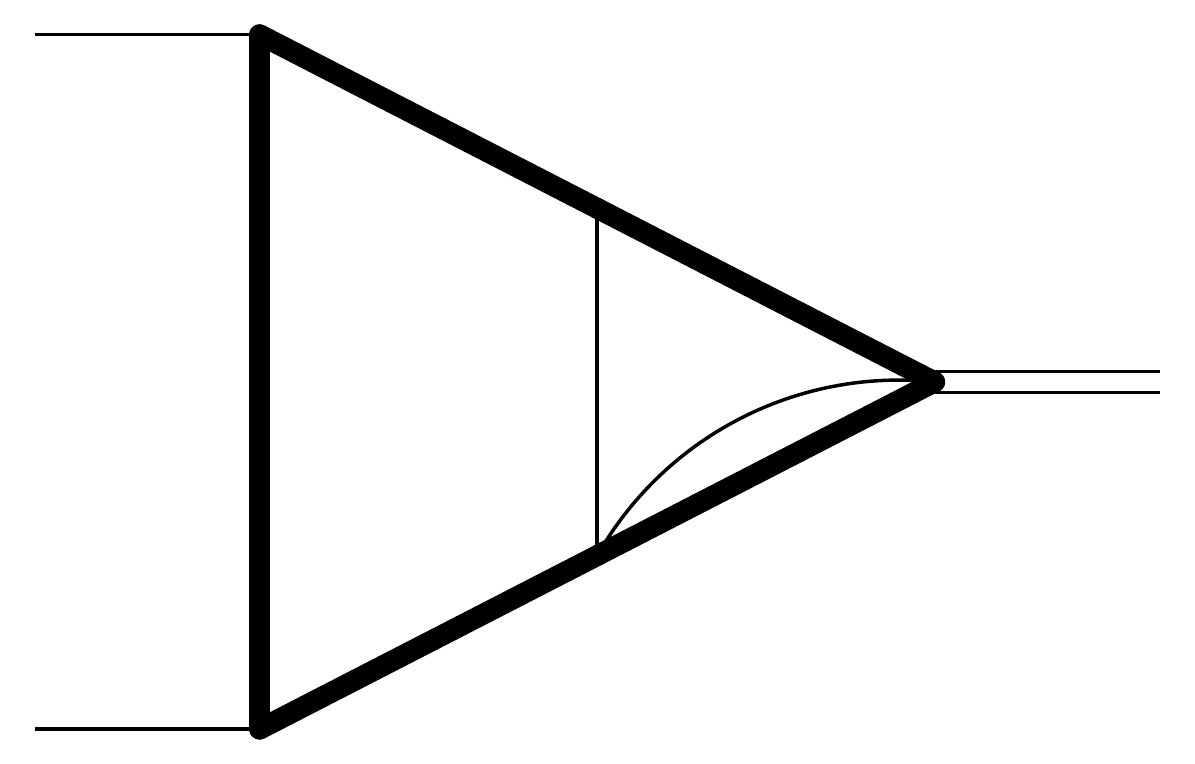}
    \caption{}
  \end{subfigure}%
  \begin{subfigure}[b]{.18\textwidth}
    \centering
    \includegraphics[scale=.2]{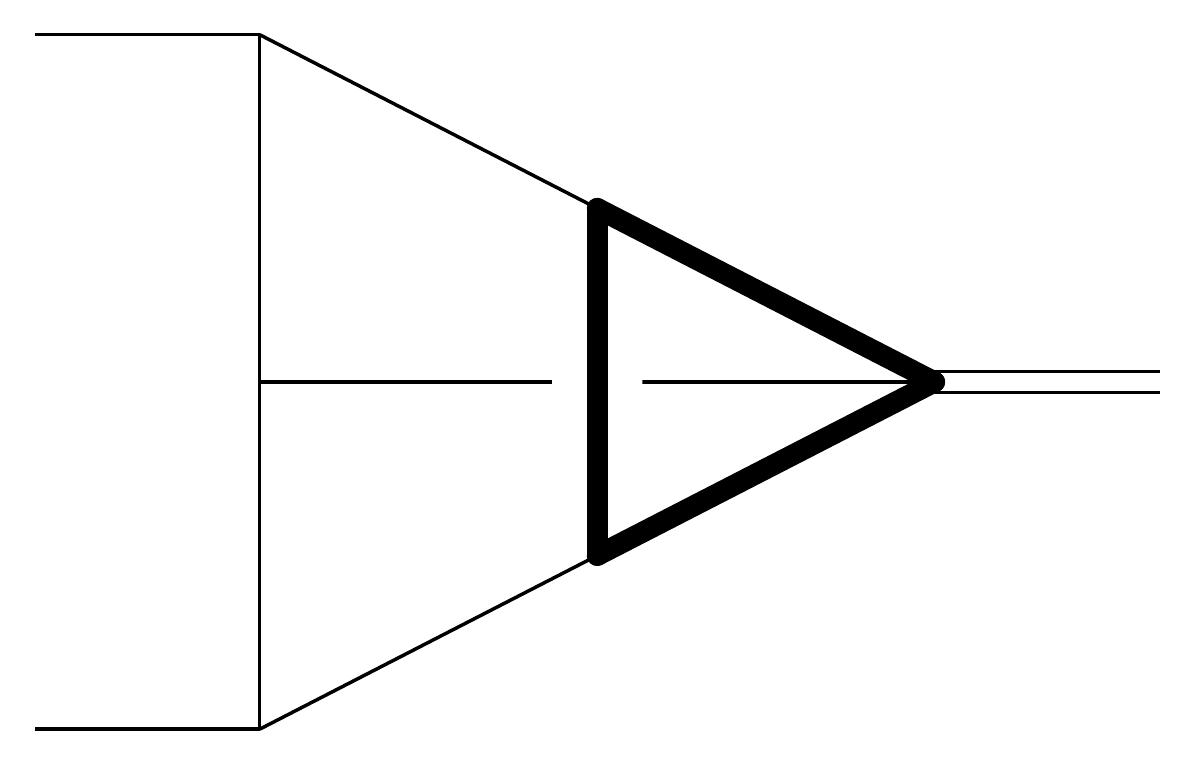}
    \caption{}
  \end{subfigure}%
  \begin{subfigure}[b]{.18\textwidth}
    \centering
    \includegraphics[scale=.2]{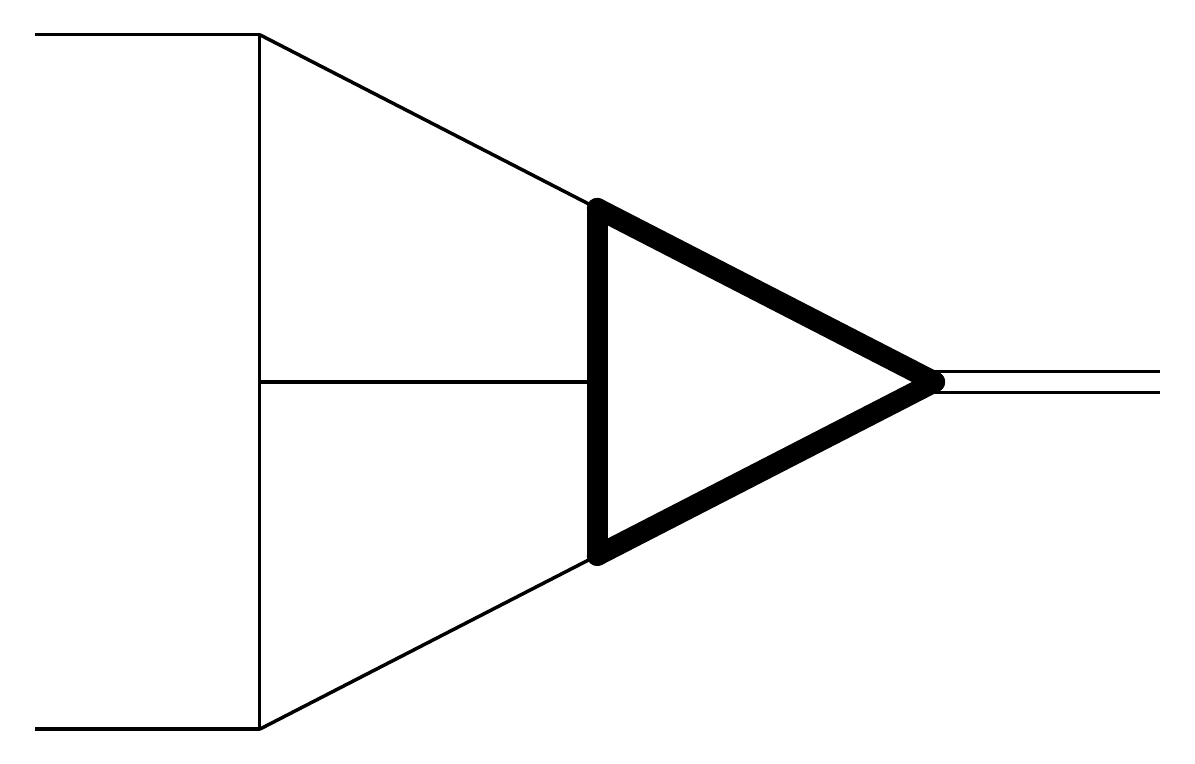}
    \caption{}
  \end{subfigure}%
  \begin{subfigure}[b]{.18\textwidth}
    \centering
    \includegraphics[scale=.2]{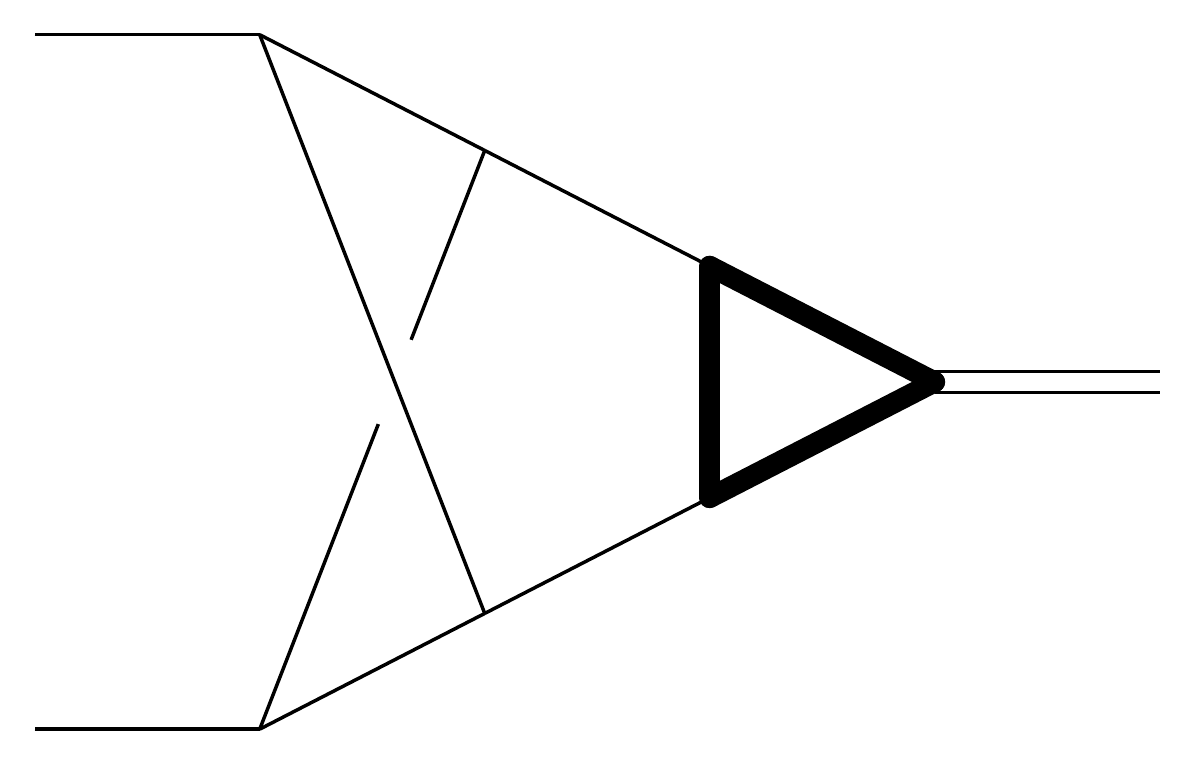}
    \caption{}
  \end{subfigure}%

  \caption{
    Sample master integrals expressible in terms of GPLs of argument $x$.
    Thin lines are massless, bold lines denote propagators with mass $\mquark$ and the outgoing double line carries the Higgs mass $\mhiggs$.
  }
  \label{fig:gpl}
\end{figure}%

178 of the 196 master integrals can be expressed completely in terms of GPLs of argument $x$.
Samples of master integrals falling in this category are shown in fig.~\ref{fig:gpl}.
The differential equations, eq.~\eqref{eq:general-de}, of these so-called canonical master integrals can be cast into an $\epsilon$-form~\cite{Henn:2013pwa}
\begin{equation}
    \frac{\d}{\d x} \vec f(x,\epsilon) = \epsilon M(x) \vec f(x,\epsilon)
\end{equation}
via a rational transformation, where the matrix $M(x)$ has only simple poles in $x$ and does not depend on $\epsilon$ anymore.
Because of the size of the problem it is quite complicated to find such a rational transformation.
Therefore, we choose a different approach, where only the homogeneous part of the differential equation is in $\epsilon$-form,
\begin{equation} \label{eq:epform-x}
    \frac{\d}{\d x} \vec f(x,\epsilon) = \epsilon M(x) \vec f(x,\epsilon) + B(x,\epsilon) \vec g(x,\epsilon)
    \,,
\end{equation}
and the matrix $B(x,\epsilon)$ has only simple poles in $x$ as well.
A basis transformation that transforms a system of differential equations of the form eq.~\eqref{eq:general-de} into the form of eq.~\eqref{eq:epform-x} can be constructed with the tool \texttt{epsilon}~\cite{Prausa:2017ltv}, which implements an algorithm presented in Ref.~\cite{Lee:2014ioa}.

The matrix $M(x)$ of eq.~\eqref{eq:epform-x} can be written as
\begin{equation} \label{eq:M-fuchsian}
    M(x)
    =
    \sum\limits_{\sigma\in S}
    \frac{M^{(\sigma)}}{x-\sigma}
    \,.
\end{equation}
The master integrals under consideration have singularities at
\begin{equation}
    S = \left\{-1,0,1,r,r^*,\varphi^2,\varphi^{-2},2,1/2\right\}\,,
\end{equation}
with a sixth root of unity $r = \mathrm{e}^{\mathrm{i}\pi/3} = (1+\mathrm{i}\sqrt3)/2$ and the golden ratio $\varphi = (1+\sqrt5)/2$.

The evaluation of the master integrals in this basis can be done in analogy to the master integrals of Ref.~\cite{Harlander:2019ioe}.
We introduce an evolution operator $U(x,x_0;\epsilon)$ which fulfills
\begin{equation}
    \frac{\d}{\d x} U(x,x_0;\epsilon)
    =
    \epsilon M(x) U(x,x_0;\epsilon)
    \quad;\quad
    U(x_0,x_0;\epsilon) = \mathds1
    \,.
\end{equation}
Making an ansatz
\begin{equation}
    U(x,x_0;\epsilon)
    =
    \sum\limits_{n=0}^\infty
    \epsilon^n
    U_n(x,x_0)
\end{equation}
leads to a recursive solution
\begin{subequations} \label{eq:U0-Un}
    \begin{align}
        U_0(x,x_0)
        &=
        \mathds1
        \,, \\
        U_{n+1}(x,x_0)
        &=
        \sum\limits_{\sigma\in S}
        \int\limits_{x_0}^x
        \frac{\d t}{t-\sigma}
        M^{(\sigma)}
        U_n(t,x_0)\,,   \label{eq:Un}
    \end{align}
\end{subequations}
where we used eq.~\eqref{eq:M-fuchsian}.
Comparing this to the definition of GPLs~\cite{Goncharov:2010jf},
\begin{equation} \label{eq:gpl-def}
    G_{\underbrace{\scriptstyle 0,\cdots,0}_{n\text{ times}}}(x)
    =
    \frac{\ln^n(x)}{n!}\,,
    \quad
    G_{a_1,\cdots,a_N}(x)
    = \int\limits_0^x \frac{\d t}{t-a_1} G_{a_2,\cdots,a_N}(t)
    \,,
\end{equation}
reveals that the integrals in eq.~\eqref{eq:Un} evaluate to GPLs of arguments $x$ and $x_0$ and indices in $S$.

The full solution of the integrals in $\vec f(x,\epsilon)$ are now given by
\begin{equation} \label{eq:f-full}
    \vec f(x,\epsilon)
    =
    \lim\limits_{x_0\rightarrow1}
    \Bigg[\,
        \int\limits_{x_0}^x \d x'\, U(x,x';\epsilon) B(x',\epsilon) \vec g(x',\epsilon)
        +
        U(x,x_0;\epsilon) \vec f(x_0,\epsilon)
    \Bigg]
    \,,
\end{equation}
where $\vec f(x_0,\epsilon)$ are the boundary conditions around $x_0=1$, i.e. the asymptotic heavy-quark expansions of the master integrals.

Since the integrals in $\vec g(x',\epsilon)$ are expressed in terms of GPLs of argument $x'$ (without rational function prefactors) and $B(x',\epsilon)$ has only simple poles in $x'$, the integrals evaluate to linear combinations of GPLs of argument $x$ as well.
Hence, the solutions $\vec f(x,\epsilon)$ are suitable to serve as input into the inhomogeneous parts of higher canonical sectors.

\subsection{Canonical sectors in $\sqrt{x}$} \label{sect:canonical-sqrtx}
\begin{figure}
  \centering
  \begin{subfigure}[b]{.18\textwidth}
    \centering
    \includegraphics[scale=.2]{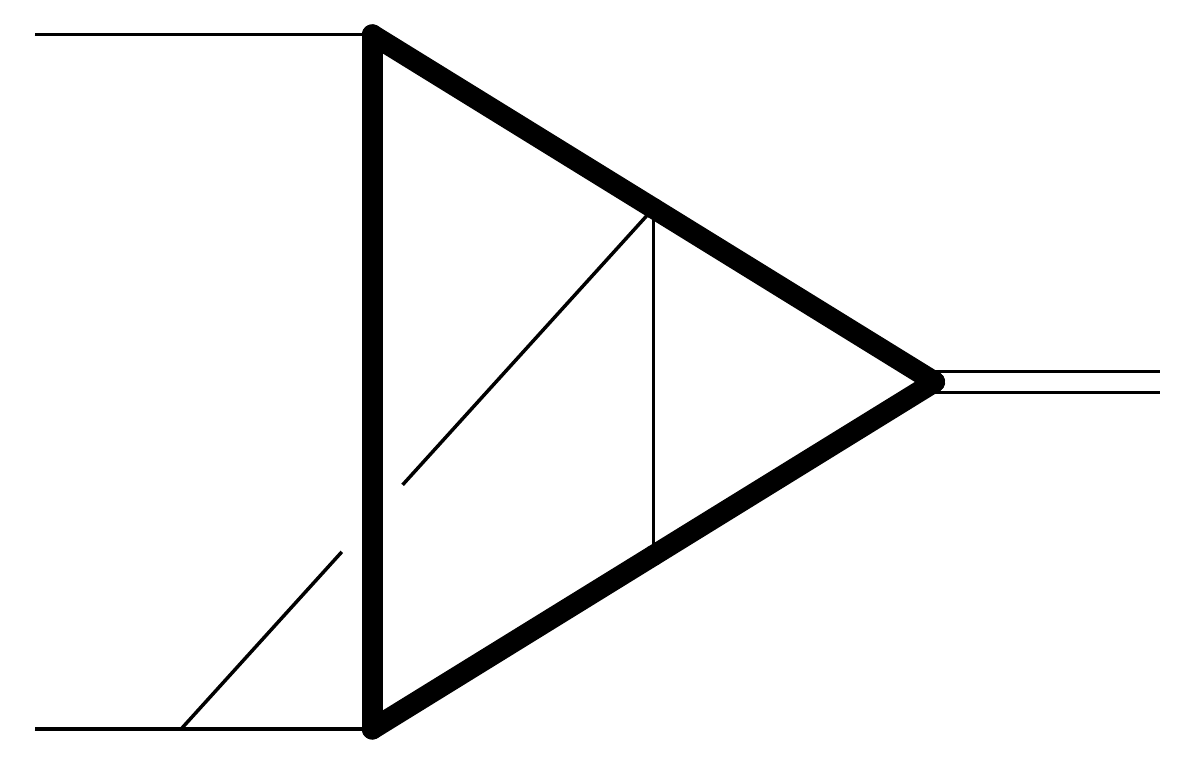}
    \caption{}
    \label{fig:sqrtx1}
  \end{subfigure}%
  \begin{subfigure}[b]{.18\textwidth}
    \centering
    \includegraphics[scale=.2]{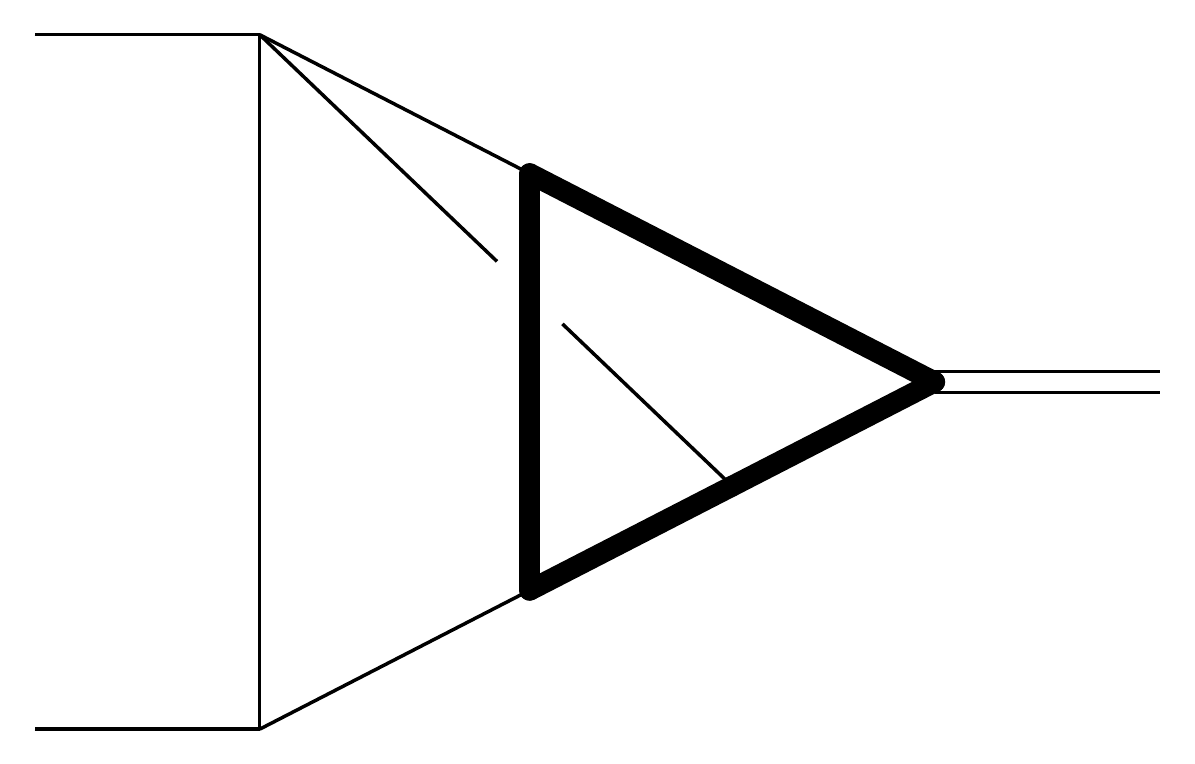}
    \caption{}
    \label{fig:sqrtx2}
  \end{subfigure}
  \caption{
    Canonical sectors in $\sqrt{x}$.
    The notation is the same as for fig.~\ref{fig:gpl}.
  }
  \label{fig:sqrtx}
\end{figure}%

The differential equations in $x$ of the two sectors shown in fig.~\ref{fig:sqrtx} cannot be transformed into $\epsilon$-form.
The sector in fig.~\ref{fig:sqrtx1} consists of three master integrals and the sector in fig.~\ref{fig:sqrtx2} of only a single master integral.
Transforming the homogeneous parts of the differential equations of those two sectors to fuchsian form using \texttt{epsilon} reveals half-integer eigenvalues of the residue matrices at $0$ and $\infty$ indicating that a change of variable is required\footnote{For a more general treatment of this situation see Ref.~\cite{Lee:2017oca}.}.

The change of variable $x = y^2$ transforms eq.~\eqref{eq:general-de} into
\begin{equation} \label{eq:de-y}
    \frac{\d}{\d y} \vec f(y,\epsilon) = 2y M(y^2,\epsilon) \vec f(y,\epsilon) + 2y B(y^2,\epsilon) \vec g(y,\epsilon)\,.
\end{equation}
This reparameterization effectively multiplies the residues at $0$ and $\infty$ by a factor of two, eliminating the half-integer eigenvalues, while all other singular points of $M(x,\epsilon)$ are split into two distinct points due to the partial fraction decomposition
\begin{equation}
    \frac1{x-\sigma}
    \rightarrow
    \frac{2y}{y^2-\sigma}
    =
    \frac1{y-\sqrt{\sigma}}
    +
    \frac1{y+\sqrt{\sigma}}
    \,.
\end{equation}
The differential equations in eq.~\ref{eq:de-y} are now in a form applicable to the tool \texttt{epsilon} in order to transform the homogenous parts into $\epsilon$-form as it was done in Section~\ref{sect:canonical-x}.

At this point, the master integrals can basically already be solved in terms of GPLs of argument $y=\sqrt{x}$ in an analogous manner to section~\ref{sect:canonical-x} but there is a technical problem.
The master integrals of lower sectors entering eq.~\ref{eq:de-y} through $\vec g(y,\epsilon)$ are expressed in terms of GPLs of argument $x$ and have to be reparameterized by the variable $y$ first.
It is straightforward to rewrite GPLs of argument $x=y^2$ into GPLs of argument $y$ by applying
\begin{equation}
    \begin{split}
        G_{a_1,\cdots,a_n}(y^2)
        &=
        \int\limits_0^{y^2}
        \frac{\d t}{t-a_1}
        G_{a_2,\cdots,a_n}(t)
        \\
        &=
        \int\limits_0^y
        \d s
        \frac{2s}{s^2-a_1}
        G_{a_2,\cdots,a_n}(s^2)
        \\
        &=
        \int\limits_0^y
        \frac{\d s}{s-\sqrt{a_1}}
        G_{a_2,\cdots,a_n}(s^2)
        +
        \int\limits_0^y
        \frac{\d s}{s+\sqrt{a_1}}
        G_{a_2,\cdots,a_n}(s^2)
    \end{split}
\end{equation}
recusively, where we used the substitution $t=s^2$ and a partial fraction decomposition.
Unfortunately, this transformation leads to a huge increase of the expression size.
Since every GPL generates $2^m$ terms, where $m$ is the number of $a_i \neq 0$, available computing resources quickly become insufficient.

A solution to this problem is to introduce a new type of multiple polylogarithms, defined by\footnote{Note that the kernel $f^0_0(t) = 2/t^2$ has a second order pole and thus should be excluded.}
\begin{subequations} \label{eq:P}
    \begin{align}
      P^{1,\cdots,1}_{\underbrace{\scriptstyle 0,\cdots,0}_{n\text{ times}}}(y)
      &=
      \frac{2^n\ln^n(y)}{n!}\,,
      \quad
      P^{n_1,\cdots,n_N}_{a_1,\cdots,a_N}(y)
      =
      \int\limits_0^y \d t\, f^{n_1}_{a_1}(t) P^{n_2,\cdots,n_N}_{a_2,\cdots,a_N}(t)
      \,,\\
      f_a^0 (t) &= \frac{2}{t^2-a}
      \,,\quad
      f_a^1 (t) = \frac{2t}{t^2-a}\,,   \label{eq:P-f}
    \end{align}
\end{subequations}
where $f_a^1(t)$ belongs to a type of integration kernels discussed for the first time in Ref.~\cite{vonManteuffel:2013vja}.
On the one hand, this definition allows rewriting every GPL in $x$ as a single $P$-function in $y$ via
\begin{equation}
    G_{a_1,\cdots,a_N}(x) = P_{a_1,\cdots,a_N}^{1,\cdots,1}(y)
    \,.
\end{equation}
On the other hand, GPL-like integration kernels appearing in eq.~\eqref{eq:de-y} can always be expressed in terms of the kernels defined in eq.~\eqref{eq:P-f} using the equation
\begin{align}
    \frac1{y-c}
    =
    \frac{c\, f_{c^2}^0(y) + f_{c^2}^1(y)}2
    \,.
\end{align}
Eq.~\eqref{eq:U0-Un} and eq.~\eqref{eq:f-full} can now be adapted to the integration kernels of eq.~\eqref{eq:P-f} and allow solving the master integrals in terms of the $P$-functions defined in eq.~\eqref{eq:P}.

In order to use existing implementations for numerical evaluation (e.g. Refs.~\cite{Vollinga:2004sn,Frellesvig:2018lmm,Naterop:2019xaf}), one can easily rewrite the $P$-functions in terms of GPLs via recursive application of
\begin{equation}
    \begin{split}
        P_{a_1,\cdots,a_N}^{0,n_2,\cdots,n_N}(y)
        &=
        \frac1{\sqrt{a_1}}
        \left[
            \int\limits_0^y \frac{\d t}{t-\sqrt{a_1}} P^{n_2,\cdots,n_N}_{a_2,\cdots,a_N}(t)
            -
            \int\limits_0^y \frac{\d t}{t+\sqrt{a_1}} P^{n_2,\cdots,n_N}_{a_2,\cdots,a_N}(t)
        \right]
        \,, \\
        P_{a_1,\cdots,a_N}^{1,n_2,\cdots,n_N}(y)
        &=
        \int\limits_0^y \frac{\d t}{t-\sqrt{a_1}} P^{n_2,\cdots,n_N}_{a_2,\cdots,a_N}(t)
        +
        \int\limits_0^y \frac{\d t}{t+\sqrt{a_1}} P^{n_2,\cdots,n_N}_{a_2,\cdots,a_N}(t)
        \,.
    \end{split}
\end{equation}
\subsection{Elliptic sector} \label{sect:elliptic}
\begin{figure}
  \centering
  \includegraphics[scale=.2]{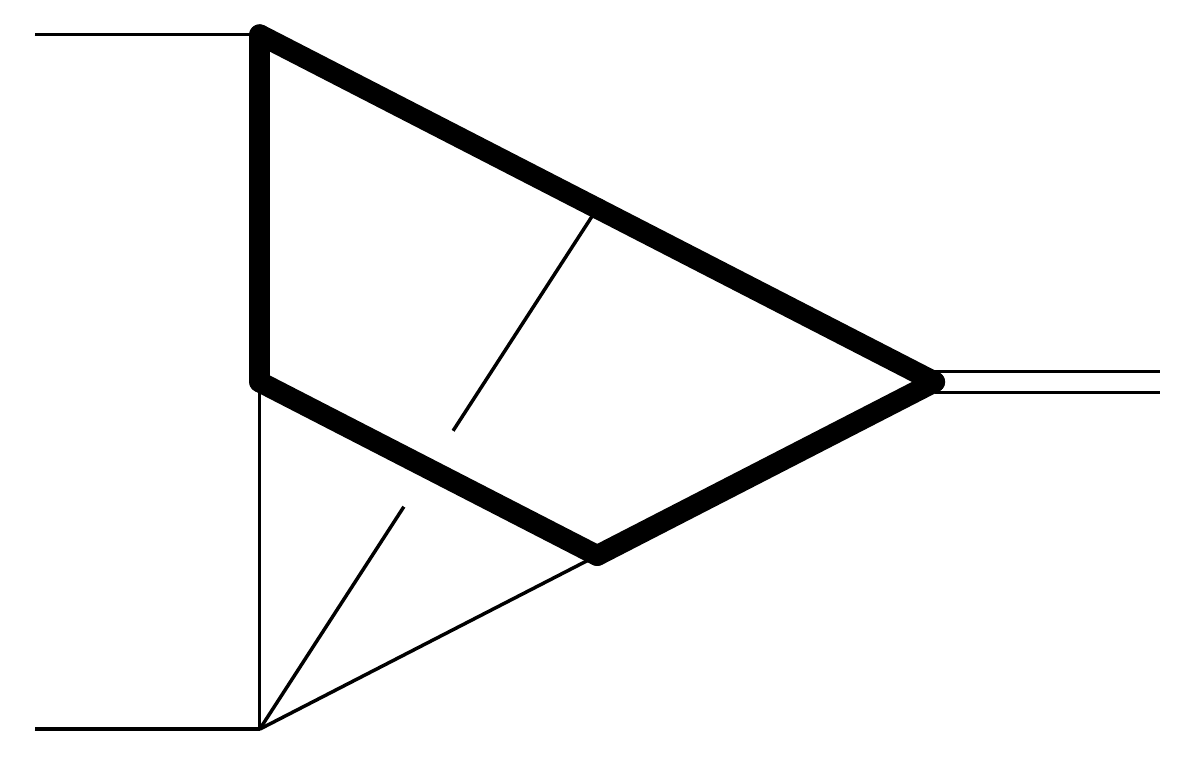}
  \caption{
    Elliptic sector.
    The notation is the same as for fig.~\ref{fig:gpl}.
  }
  \label{fig:elliptic}
\end{figure}%
The most complicated sector of the calculation is shown in fig.~\ref{fig:elliptic}.
The master integrals of this sector cannot be expressed solely in terms of GPLs but involve elliptic integrals.

As a starting point for this elliptic sector, we define the basis integrals
\begin{subequations}
    \begin{align}
        I_1(x,\epsilon)
        &=
        S_d^3
        \int\frac{\d^dl_1}{\mathrm{i}\pi^{d/2}}
        \int\frac{\d^dl_2}{\mathrm{i}\pi^{d/2}}
        \int\frac{\d^dl_3}{\mathrm{i}\pi^{d/2}}
        \frac1{D_1\cdots D_8}
        \,, \\
        I_2(x,\epsilon)
        &=
        S_d^3
        \int\frac{\d^dl_1}{\mathrm{i}\pi^{d/2}}
        \int\frac{\d^dl_2}{\mathrm{i}\pi^{d/2}}
        \int\frac{\d^dl_3}{\mathrm{i}\pi^{d/2}}
        \frac{l_2\cdot q_1}{D_1\cdots D_8}
        \,, \\
        I_3(x,\epsilon)
        &=
        S_d^3
        \int\frac{\d^dl_1}{\mathrm{i}\pi^{d/2}}
        \int\frac{\d^dl_2}{\mathrm{i}\pi^{d/2}}
        \int\frac{\d^dl_3}{\mathrm{i}\pi^{d/2}}
        \frac{l_1\cdot l_3}{D_1\cdots D_8}
        \,, \\
        I_4(x,\epsilon)
        &=
        S_d^3
        \int\frac{\d^dl_1}{\mathrm{i}\pi^{d/2}}
        \int\frac{\d^dl_2}{\mathrm{i}\pi^{d/2}}
        \int\frac{\d^dl_3}{\mathrm{i}\pi^{d/2}}
        \frac{(l_1-q_1)^2}{D_1\cdots D_8}
        \,, \\
        I_5(x,\epsilon)
        &=
        S_d^3
        \int\frac{\d^dl_1}{\mathrm{i}\pi^{d/2}}
        \int\frac{\d^dl_2}{\mathrm{i}\pi^{d/2}}
        \int\frac{\d^dl_3}{\mathrm{i}\pi^{d/2}}
        \frac{(l_1\cdot l_3)(l_2\cdot q_1)}{D_1\cdots D_8}
        \,, \\
        I_6(x,\epsilon)
        &=
        S_d^3
        \int\frac{\d^dl_1}{\mathrm{i}\pi^{d/2}}
        \int\frac{\d^dl_2}{\mathrm{i}\pi^{d/2}}
        \int\frac{\d^dl_3}{\mathrm{i}\pi^{d/2}}
        \frac{(l_1-q_1)^2(l_2\cdot q_1)}{D_1\cdots D_8}
        \,,
    \end{align}
\end{subequations}
with
\begin{equation} \label{eq:denominators}
    \begin{gathered}
        D_1 = l_1^2\,, \quad
        D_2 = (l_1 - l_2 + l_3 + q_2)^2\,, \quad
        D_3 = (l_2 - l_3 - q_1 - q_2)^2\,, \\
        D_4 = l_2^2 - 1\,, \quad
        D_5 = (l_2 - q_1 - q_2)^2 - 1\,, \quad
        D_6 = l_3^2 - 1\,, \\
        D_7 = (l_1 - l_2)^2 - 1\,, \quad
        D_8 = (l_1 - l_2 + q_2)^2 - 1\,,
    \end{gathered}
\end{equation}
and a normalization factor
\begin{equation}
    S_d = \frac{d-2}{2\Gamma(3-d/2)}\,.
\end{equation}
Since the internal mass has been set to unity in eq.~\eqref{eq:denominators}, the scalar products of the external momenta are given by
\begin{equation}
    q_1^2 = q_2^2 = 0\,, \quad
    q_1\cdot q_2 = 2z\,,
\end{equation}
with $z$ defined above eq.~\eqref{eq:variable-x}.
Even though the integrals $I_1(x,\epsilon),\cdots,I_6(x,\epsilon)$ are already finite for $\epsilon\rightarrow0$, it is more convenient to proceed in a different basis where the differential equations are fuchsian and the eigenvalues of the residues are free from resonances.
Such a basis is given by
\begin{equation}
    \begin{pmatrix} a_1(x,\epsilon) \\ \vdots \\ a_6(x,\epsilon) \end{pmatrix}
    =
    T(x,\epsilon)
    \begin{pmatrix} I_1(x,\epsilon) \\ \vdots \\ I_6(x,\epsilon) \end{pmatrix}
    \,,
\end{equation}
where
\begin{equation}
    T(x,\epsilon)
    =
    \left(
        \begin{smallmatrix}
            \frac{(1-x)^4}{2x^2} & 0 & 0 & 0 & 0 & 0 \\
            0 & -\frac{(1-x)^2}{x} & 0 & 0 & 0 & 0 \\
            -\frac{(1-22\epsilon)(1-x)^2}{x} & -4+40\epsilon & -\frac{(1-2\epsilon)(1 + x(6 + x))}{x} & \frac{2(1-2\epsilon)(1-x)^2}{x} & -4 + 8\epsilon & 4-8\epsilon \\
            0 & 0 & 0 & \frac{(1-x)^2}{2x} & 0 & 0 \\
            \frac{12(1-22\epsilon)(1-x)^2}{(1 - 2\epsilon)(2-x)} & \frac{48(1-10\epsilon)x}{(1-2\epsilon)(2-x)} & \frac{96x}{2-x} & -\frac{24(1-x)^2}{2-x} & -\frac{2 - 2(26-x)x}{2-x} & -\frac{48x}{2-x} \\
            0 & 0 & 0 & \frac{(1-x)^2(1+x^2)}{x(1 + x)} & 0 & \frac{2(1-x)^2}{1+x}
        \end{smallmatrix}
    \right)
    \,.
\end{equation}
It turns out that after expanding the functions $a_j(x,\epsilon)$ in $\epsilon$,
\begin{equation}
    a_j(x,\epsilon) = \sum_{n=0}^\infty \epsilon^n a_j^{(n)}(x)
    \,,
\end{equation}
only the seven coefficients $a_1^{(0)}(x),\cdots,a_6^{(0)}(x)$ and $a_3^{(1)}(x)$ contribute to the leading color form factor.
Since the homogeneous part of the differential equation for $a_3(x,\epsilon)$ vanishes for $\epsilon\rightarrow0$, the coefficient $a_3^{(0)}(x)$ can even be evaluated completely in terms of GPLs of argument $x$.
The six remaining coefficients, arranged in a vector $\vec a(x)$, fulfill an inhomogeneous differential equation of the form
\begin{equation} \label{eq:a-de}
    \frac{\d\vec a(x)}{\d x}
    =
    M(x)
    \vec a(x)
    +
    \vec b(x)
    \,,
\end{equation}
where the matrix $M(x)$ and the inhomogeneity $\vec b(x)$ can be found by expanding the differential equations for the $a_j(x,\epsilon)$ in $\epsilon$.

For later convenience, we introduce yet another variable $t$ by\footnote{With a slight abuse of notation, we indicate by the argument in parenthesis which of the two variables $x$ or $t$ is used to parameterize a function, i.e. $f(t) \equiv f(x=t/((4+t)(5+t)))$.}
\begin{subequations} \label{eq:variable-t}
    \begin{align}
         x &= \frac{t}{(4+t)(5+t)} \label{eq:t-to-x} \,,\\
         t &= \frac{1 - 9x - \sqrt{1 - 18x + x^2}}{2x} \label{eq:x-to-t} \,,
    \end{align}
\end{subequations}
and define once again a new basis by
\begin{subequations}
    \begin{align}
        \vec c(t)
        &=
        \left(
            c_1(t) ,\,
            \frac{\d c_1(t)}{\d t} ,\,
            \frac{\d^2 c_1(t)}{\d t^2} ,\,
            a_2^{(0)}(t) ,\,
            a_3^{(0)}(t) ,\,
            a_3^{(1)}(t)
        \right)^T
        \label{eq:vec-c}
        \,, \\
        c_1(t)
        &= \frac{t^2 + 6t + 4}{t^2 + 8t + 20}\, a_1^{(0)}(t)
        \,.
    \end{align}
\end{subequations}
After transforming the differential equation to this basis, it is straightforward to read off the third-order differential equation fulfilled by $c_1(t)$,
\begin{equation} \label{eq:de-c1}
    \begin{split}
        &\bigg[
            \frac{\d^3}{\d t^3}
            + \bigg(
                \frac3t
                + \frac3{4 + t}
                + \frac3{5 + t}
                - \frac{6(3 + t)}{4 + 6t + t^2}
            \bigg) \frac{\d^2}{\d t^2}
            \\ &\quad
            + \bigg(
                \frac1{t^2}
                - \frac{33}{5t}
                + \frac1{(4 + t)^2}
                + \frac1{4 + t}
                + \frac1{(5 + t)^2}
                - \frac{152}{5(5 + t)}
                \\ &\quad\qquad
                + \frac{120}{(4 + 6t + t^2)^2}
                + \frac{6(11 + 6t)}{4 + 6t + t^2}
                - \frac{12}{(20 + 8t + t^2)^2}
            \bigg) \frac{\d}{\d t}
            \\ &\quad
            -\frac{6}{5t^2}
            + \frac{258}{25t}
            - \frac{24}{5(5 + t)^2}
            + \frac{3648}{25(5 + t)}
            - \frac{240(3 + t)}{(4 + 6t + t^2)^3}
            \\ &\quad
            + \frac{24(-3 + 4t)}{(4 + 6t + t^2)^2}
            - \frac{9(452 + 349t)}{20(4 + 6t + t^2)}
            + \frac{24(4 + t)}{(20 + 8t + t^2)^3}
            \\ &\quad
            + \frac{3(4 + t)}{(20 + 8t + t^2)^2}
            + \frac{3(106 + 27t)}{100(20 + 8t + t^2)}
        \bigg]
        c_1(t)
        =
        k(t)
        \,.
    \end{split}
\end{equation}
Surprisingly, the inhomogeneity $k(t)$ consists of harmonic polylogarithms (HPLs) of $x$ (i.e. GPLs with indices from the set $\{-1,0,1\}$) with rational prefactors in $t$ only, despite the fact that $\vec b(x)$ in eq.~\eqref{eq:a-de} contains GPLs with indices from a larger set.
Furthermore, the diagonal block of the system of differential equations spanned by the lower three entries of $\vec c(t)$, i.e. $a_2^{(0)}(t)$, $a_3^{(0)}(t)$ and $a_3^{(1)}(t)$, vanishes completely which makes it almost trivial to solve for these coefficients after $c_1(t)$ has been found.

In Section~\ref{sect:homo-sol} we describe how to solve the homogeneous part of eq.~\eqref{eq:de-c1}, while in Section~\ref{sect:inhomo} we include the inhomogeneity $k(t)$ into the calculation which requires the introduction of a new class of iterated integrals.
\subsubsection{Homogeneous solution} \label{sect:homo-sol}
As a first step in order to evaluate the master integrals of the elliptic sector, the homogenous differential equation, eq.~\eqref{eq:de-c1} with $k(t) = 0$, has to be solved.
Similar to the differential equation of the three-loop banana diagram (see e.g. Refs.~\cite{Bloch:2014qca,Broedel:2019kmn}), the third-order differential equation, eq.~\eqref{eq:de-c1}, is a symmetric square of a second-order differential equation.
Thus, a complete set of three independent homogeneous solutions of eq.~\eqref{eq:de-c1} is given by
\begin{equation}
    c_{1,h}^{(1)}(t) = \psi_1^2(t)
    \,, \quad
    c_{1,h}^{(2)}(t) = \psi_1(t) \psi_2(t)
    \,, \quad
    c_{1,h}^{(3)}(t) = \psi_2^2(t)
    \,,
\end{equation}
where $\psi_{1,2}(t)$ solve
\begin{equation} \label{eq:de-2ndorder}
    \begin{split}
        &\bigg[
            \frac{\d^2}{\d t^2}
            +
            \bigg(
                \frac1t
                + \frac1{4+t}
                + \frac1{5+t}
                - \frac{2(3+t)}{t^2+6t+4}
            \bigg)  \frac{\d}{\d t}
            \\ &\quad
            - \frac3{5t}
            - \frac{12}{5(5 + t)}
            + \frac{15}{(4 + 6t + t^2)^2}
            + \frac{3(2 + t)}{4 + 6t + t^2}
            - \frac3{(20 + 8t + t^2)^2}
        \bigg] \psi_{1,2}(t) = 0
        \,,
    \end{split}
\end{equation}
which can be easily checked with the \texttt{MAPLE}\footnote{Maplesoft, a division of Waterloo Maple Inc., \emph{Maple 18}} package \texttt{ReduceOrder}~\cite{vanHoeij:2007}.

The differential equation~\eqref{eq:de-2ndorder} has solutions in terms of Gauss' hypergeometric functions,
\begin{subequations} \label{eq:psi12}
    \begin{align}
        \psi_1(t)
        &=
        \sqrt{2}\pi\, \pFq21{\frac14,\frac34}{1}{\frac{t(4+t)^5}{(t^2+8t+20)(t^2+6t+4)^2}}
        \,, \label{eq:psi1} \\
        \psi_2(t)
        &=
        \mathrm{i}\pi\, \pFq21{\frac14,\frac34}{1}{1-\frac{t(4+t)^5}{(t^2+8t+20)(t^2+6t+4)^2}}
        \,, \label{eq:psi2}
    \end{align}
\end{subequations}
found with another \texttt{MAPLE} package \texttt{hypergeometricsols}~\cite{Imamoglu:2017}\footnote{More general methods to obtain the homogeneous solution are given in Refs.~\cite{Primo:2016ebd,Frellesvig:2017aai,Harley:2017qut}.}.
The solutions eq.~\eqref{eq:psi12} are related to the complete elliptic integral of the first kind $K(z)$ via
\begin{equation}
    \pFq21{\frac14,\frac34}{1}{\alpha}
    =
    \frac{2}{\pi\sqrt{1 + \sqrt\alpha}}\,K\left(\frac{2\sqrt\alpha}{1 + \sqrt\alpha}\right)
    \,,
    \quad
    K(z)
    =
    \int\limits_0^{\pi/2}
    \frac{\d\varphi}{\sqrt{1-z\sin^2(\varphi)}}
    \,.
\end{equation}
\begin{figure}
  \centering
  \includegraphics[scale=.55]{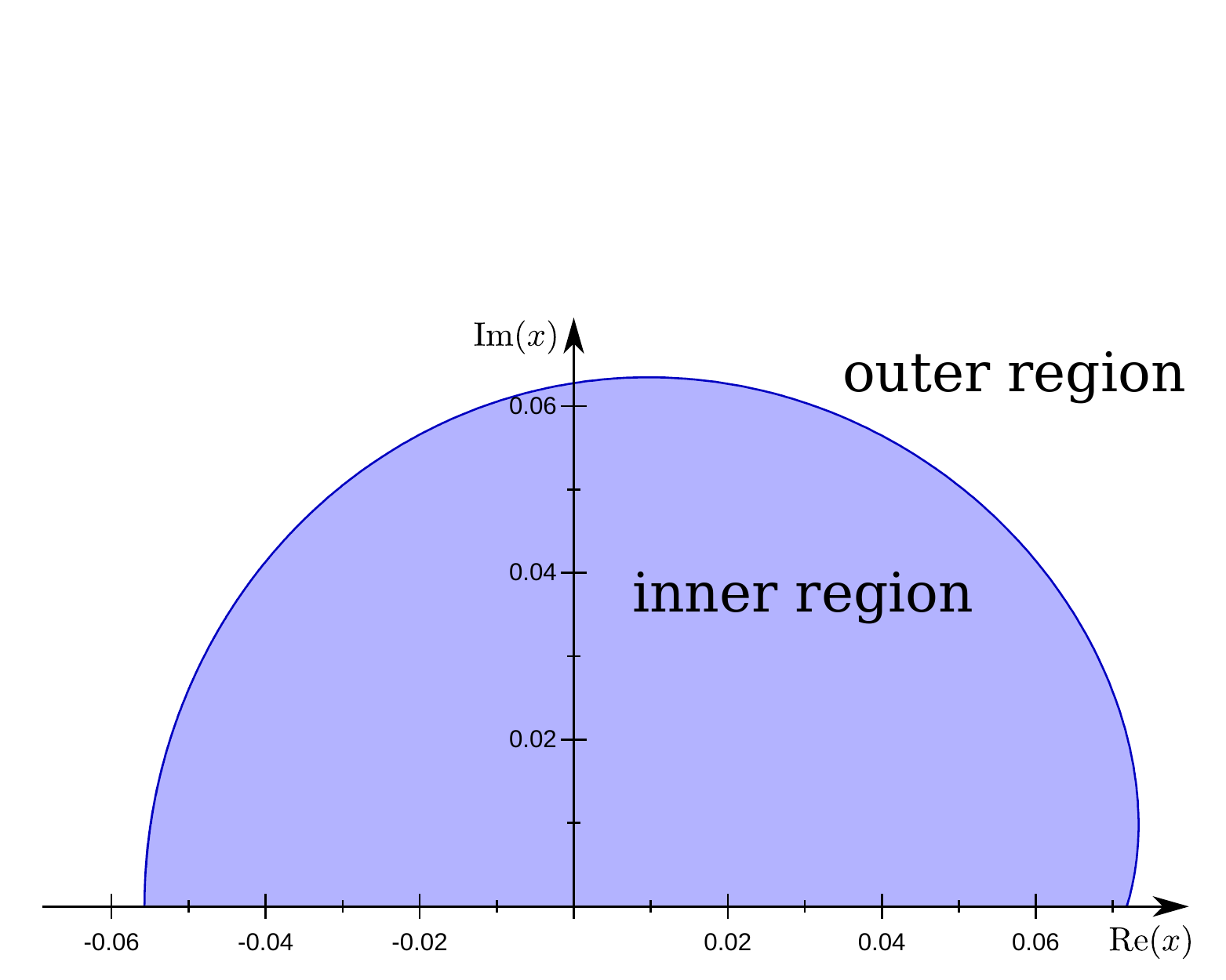}
  \caption{
    The function $\psi_1(x)$ has discontinuities for $x$ in the upper complex half-plane at the blue line which touches the real axis at $x=-9+4\sqrt{5}$ and $x=7-4\sqrt{3}$.
    In order to construct a continuous solution, one has to distinguish between the inner and outer region shown in the figure.
    \label{fig:disc}
  }
\end{figure}%
Unfortunately, the function $\psi_1(t)$ in eq.~\eqref{eq:psi1} turns out to be discontinuous at arguments that map onto the branch cut of the $\xpFq21$-function.
The position of this discontinuity (expressed through the variable $x$ related to $t$ by eq.~\eqref{eq:variable-t}) is depicted in fig.~\ref{fig:disc}, where we restrict the discussion to the physically relevant upper complex half-plane.
A continuous solution can be defined by
\begin{equation} \label{eq:psi}
    \psi(t)
    =
    \begin{cases}
        \psi_1(t) & ; \quad \text{inner region}\,, \\
        \psi_1(t) + 2\psi_2(t) & ; \quad \text{outer region}\,.
    \end{cases}
\end{equation}
An easy to implement criterion for the inner region is given by
\begin{subequations}
    \begin{equation}
        \Re(x) > -9+4\sqrt5 \;\wedge\; |x| < \frac1{10} \;\wedge\; (\Re(\alpha) \leq 1 \;\vee\; \Im(\alpha) > 0)\,,
    \end{equation}
    with
    \begin{equation}
        \alpha = \frac{t(4+t)^5}{(t^2+8t+20)(t^2+6t+4)^2}
        \,.
    \end{equation}
\end{subequations}
Furthermore, we define for later convenience
\begin{equation} \label{eq:tau}
    \uptau(t) = \frac{\psi_2(t)}{\psi(t)}
    \,.
\end{equation}
\subsubsection{Inhomogeneity} \label{sect:inhomo}
In this section, we discuss the solution to eq.~\eqref{eq:de-c1} including the inhomogeneity $k(t)$.
Since at the boundary $x_0=1$ (or equivalently $t_0=-4-2\mathrm{i}$), the function $c_1(t)$ has to vanish, the solution for $c_1(t)$ becomes
\begin{equation} \label{eq:c1-int}
    \begin{split}
        c_1(t)
        &=
        \frac{\psi^2(t)}2 \int\limits_{t_0}^t\d t\, \frac{\psi^2(t)}{w^2(t)} \uptau^2(t) \, k(t)
        -
        \uptau(t)\psi^2(t) \int\limits_{t_0}^t\d t\, \frac{\psi^2(t)}{w^2(t)} \uptau(t) \, k(t)
        \\ &\quad
        +
        \frac{\uptau^2(t)\psi^2(t)}2 \int\limits_{t_0}^t\d t\, \frac{\psi^2(t)}{w^2(t)}\, k(t)
        \,,
    \end{split}
\end{equation}
which can be obtained by a variation of constants.
The function $w(t)$ is the Wronskian of the second-order differential equation eq.~\eqref{eq:de-2ndorder} and is given by
\begin{equation}
    w(t)
    =
    \psi(t) \frac{\d\psi_2(t)}{\d t} - \psi_2(t) \frac{\d\psi(t)}{\d t}
    =
    -5\mathrm{i}\pi
    \frac{4 + 6t + t^2}{t(4 + t)(5 + t)}
    \,.
\end{equation}
When solving the integrals in eq.~\eqref{eq:c1-int}, one encounters a new class of iterated integrals, defined by
\begin{equation} \label{eq:iint}
    \IInt{f_1,\cdots,f_N}{x_0,x}
    =
    \int\limits_{[x_0]}^x
    \d x' \,
    f_1(x')\,
    \IInt{f_2,\cdots,f_N}{x_0,x'}
    \,, \quad
    \IInt{}{x_0,x} = 1
    \,.
\end{equation}
The integration kernels $f_j(x)$ that appear in this article are\footnote{Please do not confuse $\tau(x)$ in eq.~\eqref{eq:kernels:tau} with $\uptau(t)$ in eq.~\eqref{eq:tau}.}
\begin{subequations} \label{eq:kernels}
    \begin{align}
        \omega_a(x)
        &=
        \frac{1}{x-a}
        &;\quad
        a &= -1,0,1
        \label{eq:kernels:omega}
        \,,\\
        \tau(x)
        &=
        \frac{5\mathrm{i}\pi}{\psi^2(t)}
        \frac{(4 + 6t + t^2)(4 + t)(5 + t)}{t(t^2-20)}
        \label{eq:kernels:tau}
        \,,\\
        \mu_n(x)
        &=
        \frac{5\mathrm{i}\pi\psi^2(t) t^n (4 + t)(5 + t)}{t(t^2-20)(20+8t+t^2)(4+6t+t^2)}
        &;\quad
        n&=0,\cdots,6
        \label{eq:kernels:mu}
        \,,
        \\
        \kappa(x)
        &=
        \frac{5\mathrm{i}\pi\psi^2(t)\,(7t-20)(4 + t)(5 + t)}{t(t^2-20)(20+8t+t^2)^2(4+6t+t^2)}
        \label{eq:kernels:kappa}
        \,.
    \end{align}
\end{subequations}
Note that the integration in eq.~\eqref{eq:iint} is defined over the variable $x$.
The variable $t\equiv t(x)$ in eq.~\eqref{eq:kernels} is given by eq.~\eqref{eq:x-to-t}.

The notation $[x_0]$ at the lower limit of eq.~\eqref{eq:iint} indicates that a regularization procedure might be required.
In general the integral has an expansion around a point $x_1$ in powers of $\sqrt{x-x_1}$ and $\ln(x-x_1)$.
Hence, the integrals are regularized in such a way that the constant term of the expansion around $x_0$ has to vanish.
In order to not further complicate the calculation, we set $x_0=1$ for the remainder of the paper and neglect it in the notation, but we note that for $x_0=0$, this regularization method is compatible with the standard procedure of GPLs with trailing zeros.
Thus defined, the functions also obey the usual shuffle relation of iterated integrals.
Details to a numerical evaluation of this class of iterated integrals are deferred to appendix~\ref{appdx:iints}.

The integration kernels $\omega_a(x)$, eq.~\eqref{eq:kernels:omega}, are required to express the HPLs in $k(t)$ in terms of the new iterated integrals (i.e. with a lower limit $x_0=1$).
The kernel $\tau(x)$, eq.~\eqref{eq:kernels:tau}, is defined in such a way that
\begin{equation} \label{eq:tau-iint}
     \uptau(x) = \IInt{\tau}{x} + \frac{3+\mathrm{i}}{10}
     \,.
\end{equation}
The kernels $\mu_n(x)$, eq.~\eqref{eq:kernels:mu}, are modular forms for the congruence subgroup $\Gamma_0(10)$ with modular weight $4$ and constructed along the lines of Ref.~\cite{Broedel:2018rwm} with $\psi^2(t)$ as seed modular form.
Unfortunately, a double pole $(20+8t+t^2)^{-2}$ in eq.~\eqref{eq:c1-int} requires another integration kernel $\kappa(x)$, defined in eq.~\eqref{eq:kernels:kappa}, which is not a modular form.

After changing the variable in eq.~\eqref{eq:c1-int} from $t$ to $x$, one can insert eq.~\eqref{eq:tau-iint} and rewrite the HPLs in $k(t)$ in terms of the new iterated integrals (i.e. with a lower limit $x_0=1$).
The shuffle relation of iterated integrals leads then to integrals of the form of eq.~\eqref{eq:iint} with integration kernels from eq.~\eqref{eq:kernels}.

The coefficients  $a_2^{(0)}(t)$, $a_3^{(0)}(t)$ and $a_3^{(1)}(t)$ in eq.~\eqref{eq:vec-c} require an additional integration, but in doing so no further kernels have to be introduced.

\subsection{Mixed sectors} \label{sect:mixed}
\begin{figure}
  \centering
  \begin{subfigure}[b]{.18\textwidth}
    \centering
    \includegraphics[scale=.2]{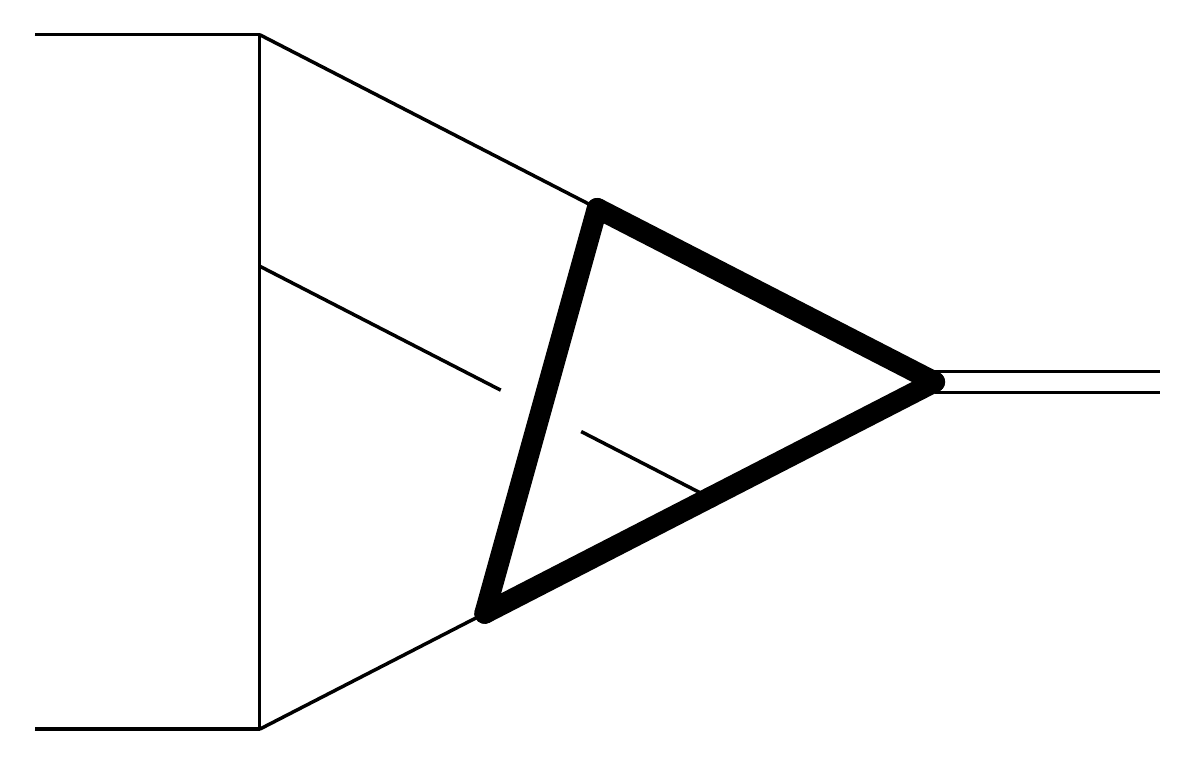}
    \caption{}
  \end{subfigure}%
  \begin{subfigure}[b]{.18\textwidth}
    \centering
    \includegraphics[scale=.2]{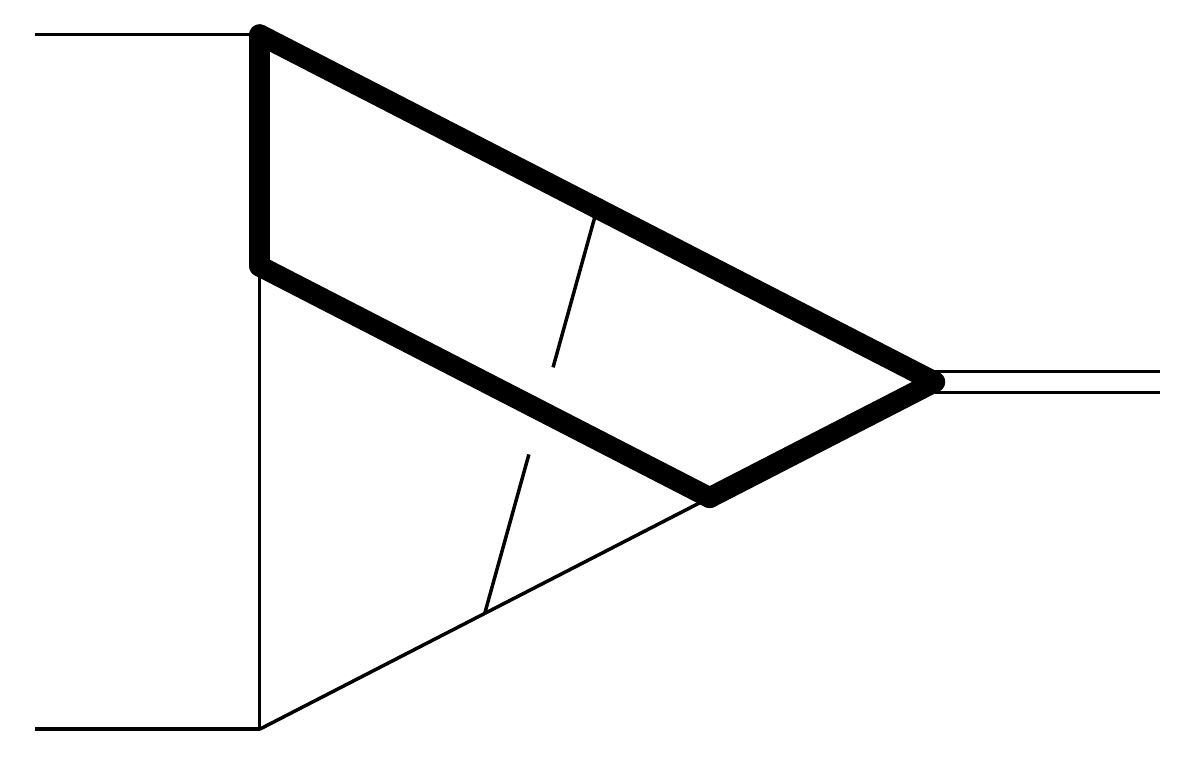}
    \caption{}
    \label{fig:mixed2}
  \end{subfigure}
  \caption{
    Mixed sectors.
    The notation is the same as for fig.~\ref{fig:gpl}.
  }
  \label{fig:mixed}
\end{figure}%
The differential equations of the sectors depicted in fig.~\ref{fig:mixed} have homogeneous parts which can be transformed into $\epsilon$-form as described in section~\ref{sect:canonical-x}.
Unfortunately, the inhomogeneities of both sectors depend on master integrals of the canonical sectors in $\sqrt{x}$ and the sector in fig.~\ref{fig:mixed2} also depends on the elliptic master integrals.

The differential equations can therefore be written as
\begin{equation}
    \begin{split}
        \frac{\d}{\d x} \vec f(x,\epsilon)
        &=
        \epsilon M(x) \vec f(x,\epsilon)
        +
        B_x(x,\epsilon) \vec g_x(x,\epsilon)
        \\ &\qquad
        +
        B_y(x,\epsilon) \vec g_y(x,\epsilon)
        +
        B_{\text{E}}(x,\epsilon) \vec g_{\text{E}}(x,\epsilon)
        \,,
    \end{split}
\end{equation}
where $\vec g_x(x,\epsilon)$ only contains the canonical master integrals in $x$ (section~\ref{sect:canonical-x}), $\vec g_y(x,\epsilon)$ the canonical master integrals in $y=\sqrt{x}$ (section~\ref{sect:canonical-sqrtx}), and $\vec g_{\text{E}}(x,\epsilon)$ the elliptic master integrals (section~\ref{sect:elliptic}).

We can now split the master integrals into three parts,
\begin{equation}
    \vec f(x,\epsilon)
    =
    \vec f_x(x,\epsilon)
    +
    \vec f_y(x,\epsilon)
    +
    \vec f_{\text{E}}(x,\epsilon)\,,
\end{equation}
so that each part obeys a differential equation where the inhomogeneity only depends on one type of master integrals, i.e.
\begin{subequations}
    \begin{align}
        \frac{\d}{\d x} \vec f_x(x,\epsilon)
        &=
        \epsilon M(x) \vec f_x(x,\epsilon)
        +
        B_x(x,\epsilon) \vec g_x(x,\epsilon)\,,    \label{eq:mixed-x}
        \\
        \frac{\d}{\d x} \vec f_y(x,\epsilon)
        &=
        \epsilon M(x) \vec f_y(x,\epsilon)
        +
        B_y(x,\epsilon) \vec g_y(x,\epsilon)\,,    \label{eq:mixed-y}
        \\
        \frac{\d}{\d x} \vec f_{\text{E}}(x,\epsilon)
        &=
        \epsilon M(x) \vec f_{\text{E}}(x,\epsilon)
        +
        B_{\text{E}}(x,\epsilon) \vec g_{\text{E}}(x,\epsilon)  \label{eq:mixed-E}
        \,.
    \end{align}
\end{subequations}
The boundary conditions could be assigned completely to $\vec f_x(x,\epsilon)$,
\begin{equation}
    \vec f_x(1,\epsilon) = \vec f(1,\epsilon)\,, \quad
    \vec f_y(1,\epsilon) = 0\,, \quad
    \vec f_{\text{E}}(1,\epsilon) = 0\,.
\end{equation}
This setup allows to solve eq.~\eqref{eq:mixed-x} as described in section~\ref{sect:canonical-x}, eq.~\eqref{eq:mixed-y} as in section~\ref{sect:canonical-sqrtx} and eq.~\eqref{eq:mixed-E} as in section~\ref{sect:elliptic}.

We want to point out, that solving for $\vec f_{\text{E}}(x,\epsilon)$ in terms of the iterated integrals, eq.~\eqref{eq:iint}, does not lead to additional integration kernels besides the ones defined in eq.~\eqref{eq:kernels}.

\section{Result} \label{sect:result}
In this section, we fix the notation of the final result and present it in numerical form.
For the sake of legibility, the lengthy analytic result has been deferred to appendix~\ref{appdx:result}.
The calculation has been done for a $SU(N_c)$ gauge group; QCD is reached for $N_c=3$.

The form factor $C$, defined in eq.~\eqref{eq:projector}, has a power series expansion in the $\MSbar$ renormalized strong coupling constant $\alpha_s$,
\begin{equation} \label{eq:result:C}
    C
    =
    \frac1v
    \frac{\alpha_s}{\pi}
    \left[
        C_0
        +
        \frac{\alpha_s}\pi
        C_1
        +
        \left(\frac{\alpha_s}\pi\right)^2
        C_2
        +
        \order{\alpha_s^3}
    \right]\,,
\end{equation}
with the vacuum expectation value $v$.
The quark mass has been renormalized in the on-shell scheme.

Starting at two-loop, the contributions are still infrared divergent.
The universal structure of these divergencies can be subtracted via~\cite{Catani:1998bh}
\begin{subequations} \label{eq:result:irsub}
  \begin{align}
    \tilde C_0
    &=
    C_0
    \,,
    \\
    \tilde C_1
    &=
    C_1 - \frac12 I_g^{(1)} C_0
    \,,
    \\
    \tilde C_2
    &=
    C_2 - \frac12 I_g^{(1)} C_1 - \frac14 I_g^{(2)} C_0
    \,,
    \label{eq:IRsubtract}
  \end{align}
\end{subequations}
where the factors $I_g^{(1)}$ and $I_g^{(2)}$ are given by~\cite{Catani:1998bh,deFlorian:2012za}
\begin{subequations}
  \begin{align}
    I_g^{(1)}
    &\equiv
    I_g^{(1)}(\epsilon)
    =
    -\left(-\frac{\mu^2}{\mhiggs^2}\right)^{\epsilon}
    \frac{e^{\epsilon\gamma_E}}{\Gamma(1-\epsilon)}
    \left[\frac{N_c}{\epsilon^2} + \frac{\beta_0}{\epsilon}\right]
    \,, \\
    \begin{split}
      I_g^{(2)}
      &=
      -
      \frac12 I_g^{(1)}(\epsilon)
      \left(
        I_g^{(1)}(\epsilon)
        +
        \frac{\beta_0}{\epsilon}
      \right)
      +
      \frac{e^{-\epsilon\gamma_E} \Gamma(1-2\epsilon)}{\Gamma(1-\epsilon)}
      \left(
        \frac{\beta_0}{\epsilon}
        +
        K
      \right)
      I_g^{(1)}(2\epsilon)
      \\ &\qquad
      +
      \left(-\frac{\mu^2}{\mhiggs^2}\right)^{2\epsilon}
      \frac{e^{\epsilon\gamma_E}}{\Gamma(1-\epsilon)}
      \frac{H_g}{2\epsilon}
      \,,
    \end{split}
  \end{align}
\end{subequations}
and
\begin{subequations}
  \begin{align}
    \beta_0
    &=
    \frac{11}6 N_c
    -
    \frac13 n_l
    \,, \\
    K
    &=
    \left(\frac{67}{18} - \frac{\pi^2}6\right) N_c
    -
    \frac{5}{9} n_l
    \,, \\
    \begin{split}
        H_g
        &=
        \frac{5}{27} n_l^2
        +
        \frac14
        \left(N_c-\frac1{N_c}\right)
        n_l
        -
        \left(\frac{\pi^2}{72} + \frac{29}{27}\right) n_l N_c
        \\ &\quad
        +
        \left(\frac{\zeta_3}2 + \frac5{12} + \frac{11\pi^2}{144}\right) N_c^2
        \,.
    \end{split}
  \end{align}
\end{subequations}
We expand the finite contributions $\tilde C_i$ in powers of $N_c$ as
\begin{equation} \label{eq:result:CiNc}
    \tilde C_0
    =
    \tilde C_0^{(0)}
    \,, \quad
    \tilde C_1
    =
    \sum_{k=-1}^1
    N_c^k
    \,
    \tilde C_1^{(k)}
    \,, \quad
    \tilde C_2
    =
    \sum_{k=-2}^2
    N_c^k
    \,
    \tilde C_2^{(k)}
    \,.
\end{equation}
In order to fix the notation, we provide the one-loop result,
\begin{equation} \label{eq:result:1loop}
    \tilde C_0^{(0)}
    =
    -\frac{2x}{(x-1)^2}
    +
    \frac{x(x+1)^2}{(x-1)^4} G_{0,0}(x)
    \,.
\end{equation}
The leading color contribution $\tilde C_2^{(2)}$ is shown in fig.~\ref{fig:result}.
The analytic result will be presented, along with the two-loop results, in appendix~\ref{appdx:result} and in electronic form in a supplementary file, see appendix~\ref{appdx:suppl:result}.
\begin{figure}
    \centering
    \includegraphics{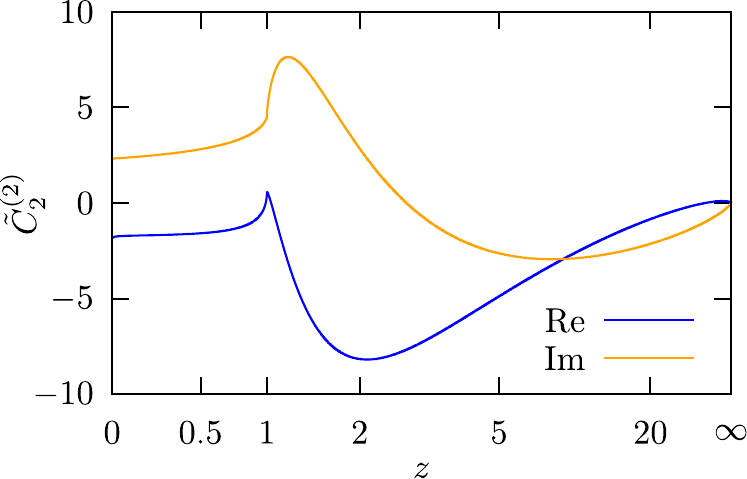}
    \caption{
        Real and imaginary part of the three-loop leading color contribution $\tilde C^{(2)}_2$ to the \ggh form factor.
        The renormalization scale has been set to $\mu^2=\mhiggs^2$.
        \label{fig:result}
    }
\end{figure}

We checked analytically that our result agrees with the heavy-quark expansion of Refs.~\cite{Davies:2019nhm,Harlander:2009mq,Pak:2009bx} up to ${\cal O}((1-x)^{14})$.
The non-analytic terms of the threshold expansion, provided in Ref.~\cite{Grober:2017uho}, could also be reproduced.

For $x=−0.00112+\mathrm{i}0$, corresponding approximately to a standard model bottom quark of mass $\mquark = \mbottom=4.18\,\mathrm{GeV}$ and an on-shell Higgs-boson of mass $\mhiggs=125.18\,\mathrm{GeV}$~\cite{Tanabashi:2018oca}, we obtain a value of
\begin{equation}
    \left. \tilde C^{(2)}_2 \right|_{\text{bottom}} = 0.10973435 -  0.40966435\,\mathrm{i}\,.
\end{equation}
This is of course only a partial result not including contributions of any other quarks or sub-leading color terms.

\section{Conclusions and outlook} \label{sect:conclusion}
In this paper, we presented for the first time an analytic result for the leading color contribution to the three-loop \ggh form factor in QCD with a finite mass of the mediating quark.
Previously at three-loop, analytic expressions were only available for contributions involving one light-fermion loop~\cite{Harlander:2019ioe}.
A full result that includes also sub-leading color terms is still only known in numerical form~\cite{Czakon:2020vql}.

We showed that our result requires the introduction of a new class of iterated integrals, eqs.~(\ref{eq:iint},~\ref{eq:kernels}), with integration kernels involving elliptic integrals.
These iterated integrals can be evaluated numerically to high precision, see Appendix~\ref{appdx:iints}.

We note that the iterated integrals of this article are not iterated integrals of modular forms studied in Refs.~\cite{Manin:2006,Brown:2014,Adams:2017ejb,Broedel:2018iwv,Adams:2018ulb,Broedel:2018rwm,Duhr:2019rrs}.
Iterated integrals of modular forms are iterated integrals in the variable $\uptau$, defined in eq.~\eqref{eq:tau}, where the kernels are modular forms for a congruence subgroup of the modular group $\mathrm{SL}_2(\mathds{Z})$.
The kernels $\mu_n(x) \d x/\d\uptau$, eq.~\eqref{eq:kernels:mu}, are indeed modular forms for the congruence subgroup $\Gamma_0(10)$ with modular weight $4$.
Likewise, the kernel $\tau(x) \d x/\d\uptau=1$, eq.~\eqref{eq:kernels:tau}, is a modular form of weight $0$ and $\omega_0(x) \d x/\d\uptau$, eq.~\eqref{eq:kernels:omega} for $a=0$, is a modular form of weight $2$.
Unfortunately, the remaining kernels, $\omega_{\pm1}(x) \d x/\d\uptau$, eq.~\eqref{eq:kernels:omega} for $a=\pm1$, and $\kappa(x) \d x/\d\uptau$, eq.~\eqref{eq:kernels:kappa}, are not modular forms.
We did not find any obvious linear relation between the iterated integrals used in this article, but in order to exclude the existence of such relations completely a thorough analysis is still required.
It might still possible that the result can be expressed completely in terms of iterated integrals of modular forms after all.
Another interesting topic would be to analyse if our result can be expressed in terms of the elliptic polylogarithms known in the literature~\cite{Adams:2013nia,Adams:2014vja,Adams:2015gva,Adams:2015ydq,Adams:2016vdo,Adams:2016xah,Broedel:2017kkb,Broedel:2017siw,Broedel:2018qkq,Broedel:2019hyg}.

For the sub-leading color contributions, we expect more elliptic sectors to appear.
If these sectors have homogeneous solutions that can be solved as easily as the sector discussed in this paper, it should be possible to provide analytic results for these contributions as well.
This would most certainly require the introduction of a larger set of integration kernels for the iterated integrals.

\section*{Acknowledgments}
We would like to thank Robert~Harlander for useful discussions and his comments on the manuscript.
J.U.\ received funding in the early stage of this work from the European Research Council (ERC) under the European Union’s Horizon 2020 research and innovation programme under grant agreement no.\ 647356 (CutLoops).
The authors acknowledge support by the state of Baden-Württemberg through bwHPC and the German Research Foundation (DFG) through grant no INST 39/963-1 FUGG.
Parts of the computing resources were granted by RWTH Aachen University under project rwth0119.
We would also like to thank Peter Uwer and his group ``Phenomenology of Elementary Particle Physics beyond the Standard Model'' at Humboldt-Universit\"at zu Berlin for providing computer resources.

\appendix
\section{Numerical evaluation of iterated integrals} \label{appdx:iints}
In this appendix, we discuss the numerical evaluation of the iterated integrals defined in eqs.~(\ref{eq:iint},~\ref{eq:kernels}).

It is possible to derive a recursive algorithm for a series expansions for an iterated integral around an arbitrary point $x_1$ in powers of $\sqrt{x-x_1}$ and $\ln(x-x_1)$, where almost all coefficients are expressed through coefficients of an iterated integral with one integration less.
Only the constant term in this series expansion cannot be determined that way.

The strategy for a numerical evaluation is to start at $x_0$, where the constant terms vanish by definition, and fit the constant term of an expansion around a point $x_1$, not too far away, by evaluating both series at a point in between that lies within the convergence radii.
By repeating this procedure for points along a path, every point of interest can be reached.

Section~\ref{appdx:iints:iint-series} explains the basics for a series expansion of iterated integrals.
Section~\ref{appdx:iints:kernel-series} describes the expansion of the integration kernels in eq.~\eqref{eq:kernels}.

A \texttt{C++} implementation is included in the supplementary material and will be discussed in Section~\ref{appdx:suppl:libiint}.

\subsection{Series expansion of iterated integrals} \label{appdx:iints:iint-series}
As an ansatz, we assume that eq.~\eqref{eq:iint} has a series expansion around $x=x_1$ of the form
\begin{equation} \label{eq:iint-series}
    \IInt{f_1,\cdots,f_N}{x_0,x_1+\delta}
    =
    \sum\limits_{n=0}^\infty
    \sum\limits_{m=0}^\infty
    c^{(n,m)}_{f_1,\cdots,f_N}(x_0,x_1)
    \delta^{n/2}
    \ln^m(\delta)
    \,,
\end{equation}
while the integration kernels expand to
\begin{equation} \label{eq:kernel-series}
    f_j(x_1+\delta)
    =
    \sum\limits_{k=k_0}^\infty
    a_{f_j}^{(k)}(x_1)
    \delta^{k/2}
    \,.
\end{equation}
Shifting the integration variable in eq.~\eqref{eq:iint} yields
\begin{equation}
    \IInt{f_1,\cdots,f_N}{x_0,x_1+\delta}
    =
    \hspace{-10pt}
    \int\limits_{[x_0-x_1]}^\delta
    \hspace{-10pt}
    \d\delta' \,
    f_1(x_1+\delta')\,
    \IInt{f_2,\cdots,f_N}{x_0,x_1+\delta'}
    \,.
\end{equation}
Inserting eqs.~(\ref{eq:iint-series},~\ref{eq:kernel-series}) leads to
\begin{equation} \label{eq:iint-series-inserted}
    \begin{split}
        \IInt{f_1,\cdots,f_N}{x_0,x_1+\delta}
        &=
        \sum\limits_{n=k_0}^\infty
        \sum\limits_{k=k_0}^n
        \sum\limits_{m=0}^\infty
        a_{f_1}^{(k)}(x_1)
        c^{(n-k,m)}_{f_2,\cdots,f_N}(x_0,x_1)
        \\ &\qquad\times
        \hspace{-10pt}
        \int\limits_{[x_0-x_1]}^\delta
        \hspace{-10pt}
        \d\delta' \,
        \delta'^{n/2}
        \ln^m(\delta)
        \,.
    \end{split}
\end{equation}
The integral is solved by
\begin{equation} \label{eq:aux-int}
    \begin{split}
        \int\limits^\delta \d\delta'\,
        \delta'^{n/2}
        \ln^m(\delta')
        &=
        \begin{cases}
            -m!\,
            \delta^{n/2+1}
            \sum\limits_{j=0}^m
            \frac{(-1-n/2)^{j-m-1}}{j!}
            \ln^j(\delta)
            &;\quad n\neq-2
            \\
            \frac{\ln^{m+1}(\delta)}{m+1}
            &;\quad n=-2
        \end{cases}
        \\[3pt] &\quad
        +
        \text{const.}
    \end{split}
\end{equation}
The constant in eq.~\eqref{eq:aux-int} depends on the lower limit and contributes only to the coefficient $c^{(0,0)}_{f_1,\cdots,f_N}(x_0,x_1)$ in eq.~\eqref{eq:iint-series} that cannot be expressed through lower weight coefficients, anyway.
Hence, eq.~\eqref{eq:iint-series-inserted} becomes
\begin{equation}
    \begin{split}
        &\IInt{f_1,\cdots,f_N}{x_0,x_1+\delta}
        \\
        &\qquad=
        c^{(0,0)}_{f_1,\cdots,f_N}(x_0,x_1)
        +
        \sum\limits_{m=0}^\infty
        \frac{\ln^{m+1}(\delta)}{m+1}
        \sum\limits_{k=k_0}^{-2}
        a_{f_1}^{(k)}(x_1)
        c^{(-k-2,m)}_{f_2,\cdots,f_N}(x_0,x_1)
        \\
        &\qquad\quad
        -
        \sum\limits_{\substack{n=k_0+2\\n\neq0}}^\infty
        \sum\limits_{m=0}^\infty
        \delta^{n/2}
        \ln^m(\delta)
        \sum\limits_{k=k_0}^{n-2}
        \sum\limits_{j=m}^\infty
        \frac{j!\,(-n/2)^{m-j-1}}{m!}
        a_{f_1}^{(k)}(x_1)
        c^{(n-k-2,j)}_{f_2,\cdots,f_N}(x_0,x_1)
        \,.
    \end{split}
\end{equation}
This equation allows us to read off the coefficients of eq.~\eqref{eq:iint-series},
\begin{subequations} \label{eq:coeff-rec}
    \begin{align}
        c^{(0,m>0)}_{f_1,\cdots,f_N}(x_0,x_1)
        &=
        \sum\limits_{k=k_0}^{-2}
        a_{f_1}^{(k)}(x_1)
        c^{(-k-2,m)}_{f_2,\cdots,f_N}(x_0,x_1)
        \,, \\
        c^{(n\neq 0,m)}_{f_1,\cdots,f_N}(x_0,x_1)
        &=
        -
        \sum\limits_{k=k_0}^{n-2}
        \sum\limits_{j=m}^\infty
        \frac{j!\,(-n/2)^{m-j-1}}{m!}
        a_{f_1}^{(k)}(x_1)
        c^{(n-k-2,j)}_{f_2,\cdots,f_N}(x_0,x_1)
        \,.
    \end{align}
\end{subequations}
The initial conditions for the recursion are given by the expansion coefficients of the weight zero function $\IInt{}{x_0,x} = 1$,
\begin{equation}
    c^{(n,m)}(x_0,x_1) = \begin{cases}
        1 &;\quad n=m=0 \\
        0 &;\quad \text{else}\,.
    \end{cases}
\end{equation}
\subsection{Series expansion of integration kernels} \label{appdx:iints:kernel-series}
Since the expansion coefficients of the integration kernels, eq.~\eqref{eq:kernels}, enter eq.~\eqref{eq:coeff-rec}, it is disirable to find efficient approaches to determine these quantities.
The GPL-like kernels, eq.~\eqref{eq:kernels:omega}, are trivial to expand.
Unfortunately, this is not true for all the other integration kernels as they involve the squared homogeneous solution $\psi^2(t)$ given in eq.~\eqref{eq:psi}.

Instead of $\psi^2(t)$, we consider the function
\begin{equation}
    \phi(t)
    =
    \frac{20 + 8t + t^2}{4 + 6t + t^2} \psi^2(t)
    \,,
\end{equation}
which fulfills a homogeneous third-order differential equation in the variable $x$,
\begin{equation}
    \begin{split}
        &\bigg[
            2(1 + x)(1 + 28x + x^2)
            +
            (-1 + x)(1 - 56x - 46x^2 - 8x^3 + x^4)\frac{\d}{\d x}
            \\ &\qquad
            +
            3x(-1 + x)^2(-1 + 27x - 11x^2 + x^3) \frac{\d^2}{\d x^2}
            \\ &\qquad
            +
            x^2(-1 + x)^3(1 - 18x + x^2) \frac{\d^3}{\d x^3}
        \bigg]
        \phi(x)
        =
        0
        \,.
    \end{split}
\end{equation}
This differential equation can be solved with a power series ansatz,
\begin{equation} \label{eq:phi-ansatz}
    \phi(x_1+\delta)
    =
    \sum_{k=k_0}^\infty
    b^{(k)}(x_1)
    \delta^{k/2}\,,
\end{equation}
that is sufficient for all points $x_1$.

Furthermore, the integration kernels involve rational functions in $t$, which have to be expanded in $x$.
We perform these expansions in three seperate steps.
The expansion of such a function in $t$ as well as the expansion of $t$ in $x$ can be performed in a straightforward manner.
In a third step, these expansions have to be composed using Fa\`{a} di Bruno's formula~\cite{Bruno:1855}.

Via the Cauchy product, the expansion of $\phi(x)$ can be combined with the $x$-expansion of a rational function in $t$ in order to form the integration kernels in eq.~\eqref{eq:kernels:kappa} and eq.~\eqref{eq:kernels:mu}.
The kernel $\tau(x)$, eq.~\eqref{eq:kernels:tau}, requires in addition the construction of a reciprocal power series.

\vspace{-.5em}
\section{Supplementary material}
As supplementary material to this article, we provide the main result in electronic form, see section~\ref{appdx:suppl:result}.
In addition, we present in section~\ref{appdx:suppl:libiint} an implementation of the algorithm discussed in appendix~\ref{appdx:iints} for a numerical evaluation of the iterated integrals defined in eqs.~(\ref{eq:iint},~\ref{eq:kernels}).

\subsection{\texttt{ggh-nc2.m}} \label{appdx:suppl:result}
The supplementary file \texttt{ggh-nc2.m} contains the results of this article in a \texttt{Mathematica}\footnote{Wolfram Research, Inc., \emph{Mathematica, Version 12.0}, Champaign, IL, U.S.A.} readable format using the following notation:
\begin{center}
    \begin{longtable}{@{}p{6.5cm}p{3.5cm}p{5cm}@{}}
        \toprule
        \texttt{Cggh} & $\tilde C$ & -- \\
        \texttt{cggh\lbracket}$k$\texttt{\rbracket} & $\tilde C_k$ & eqs.~(\ref{eq:result:irsub},~\ref{eq:result:CiNc}) \\
        \texttt{cggh[0,0]} & $\tilde C_0^{(0)}$ & eq.~\eqref{eq:result:1loop} \\
        \texttt{cggh[1,-1]} & $\tilde C_1^{(\shortminus1)}$ & eq.~\eqref{eq:result:C1-1} \\
        \texttt{cggh[1,0]} & $\tilde C_1^{(0)}$ & eq.~\eqref{eq:result:C10} \\
        \texttt{cggh[1,1]} & $\tilde C_1^{(1)}$ & eq.~\eqref{eq:result:C11} \\
        \texttt{cggh[2,2]} & $\tilde C_2^{(2)}$ & eq.~\eqref{eq:main-result} \\
        \midrule
        \texttt{aspi} & $\alpha_s/\pi$ & -- \\
        \texttt{Lmu} & $L_\mu$ & appendix~\ref{appdx:result} \\
        \texttt{phi} & $\varphi = (1+\sqrt5)/2$ & golden ratio \\
        \texttt{r1} & $r = \mathrm{e}^{\mathrm{i}\pi/3}$ & sixth root of unity \\
        \texttt{r2} & $r^* = \mathrm{e}^{-\mathrm{i}\pi/3}$ & sixth root of unity \\
        \texttt{GPL\lbracket\ltbrace}$a_1$\texttt{,}$\cdots$\texttt{,}$a_N$\texttt{\rtbrace,x\rbracket} & $G_{a_1,\cdots,a_N}(x)$ & eq.~\eqref{eq:gpl-def} \\
        \texttt{Plog\lbracket\ltbrace\ltbrace}$n_1$\texttt{,}$a_1$\texttt{\rtbrace,}$\cdots$\texttt{,\ltbrace}$n_N$\texttt{,}$a_N$\texttt{\rtbrace\rtbrace,y\rbracket} & $P^{n_1,\cdots,n_N}_{a_1,\cdots,a_N}(y)$ & eq.~\eqref{eq:P} \\
        \texttt{IInt\lbracket\ltbrace}$f_1$\texttt{,}$\cdots$\texttt{,}$f_N$\texttt{\rtbrace,x\rbracket} & $\IInt{f_1,\cdots,f_N}{1,x}$ & eq.~\eqref{eq:iint}, appendix~\ref{appdx:suppl:libiint:mma} \\
        \texttt{psi} & $\psi(t)$ & eq.~\eqref{eq:psi} \\
        \texttt{dpsi} & $\frac{\d\psi(t)}{\d t}$ & -- \\
        \texttt{d2psi} & $\frac{\d^2\psi(t)}{\d t^2}$ & -- \\
        \bottomrule
    \end{longtable}
\end{center}

\subsection{\texttt{libiint}} \label{appdx:suppl:libiint}
In this section, we present the \texttt{C++} library \texttt{libiint}\footnote{licensed under the \emph{GNU Lesser General Public License, version 3}.} to evaluate the iterated integrals of eqs.~(\ref{eq:iint},~\ref{eq:kernels}).
The library is easily accessible through an included \texttt{Mathematica} interface.
All necessary files are provided as supplementary material to the published version of this article or can be obtained from \texttt{GitHub} using
\begin{quote}
    \texttt{git clone https://github.com/mprausa/libiint.git}
\end{quote}

Section~\ref{appdx:suppl:libiint:install} explains the installation procedure and section~\ref{appdx:suppl:libiint:mma} the \texttt{Mathematica} interface.
\subsubsection{Installation} \label{appdx:suppl:libiint:install}
The \texttt{libiint} package provides a \texttt{CMake}\footnote{\url{https://www.cmake.org}} build system to compile and install the library and the \texttt{Mathematica} interface.

The following dependencies have to be properly installed before the compilation can be started:
\begin{itemize}
    \item \texttt{FLINT: Fast Library for Number Theory}\footnote{\url{http://www.flintlib.org}},
    \item \texttt{Arb - a C library for arbitrary-precision ball arithmetic}\footnote{\url{http://www.arblib.org}},
    \item \texttt{yaml-cpp: A YAML parser and emitter in C++}\footnote{\url{https://github.com/jbeder/yaml-cpp}}.
\end{itemize}
Obviously, \texttt{Wolfram Mathematica} is required in addition to build the \texttt{Mathematica} interface.

Within the \texttt{libiint/} directory, we recommend to create a new directory \texttt{build/} and change into it.
Now inside the \texttt{build/} directory run \texttt{CMake} with
\begin{quote}
    \texttt{cmake }\textit{[options]}\texttt{ ..}
\end{quote}
where valid options are
\begin{itemize}
    \item \texttt{-DCMAKE\_INSTALL\_PREFIX=/path/to/install} to install the package at \\ \texttt{/path/to/install},
    \item \texttt{-DBUILD\_MMA\_INTERFACE=OFF} to disable building the \texttt{Mathematica} interface,
    \item \texttt{-DINSTALL\_MMA\_DIR=/path/to/install} to specify the location where the \\ \texttt{Mathematica} package will be installed. \\ (default: \texttt{\$UserBaseDirectory/Applications/IInt})
\end{itemize}
The package can now be compiled and installed via
\begin{quote}
    \texttt{make} \\
    \texttt{make install}
\end{quote}
where the second line may require \text{root} privileges depending on the installation paths.
You might also have to adjust \texttt{Mathematica}'s \texttt{\$Path} variable so that it includes the installation directory of the \texttt{Mathematica} interface.

\subsubsection{\texttt{Mathematica} interface} \label{appdx:suppl:libiint:mma}
The library \texttt{libiint} obtains its usefulness only through the \texttt{Mathematica} interface, which is why we refrain from discussing direct access to the library in this article.

After a successfull installation, you should be able to load the package via
\begin{quote}
    \texttt{<<\,IInt\textasciigrave}
\end{quote}
from within \texttt{Mathematica} without getting any errors.

The iterated integrals, eq.~\eqref{eq:iint}, are represented by
\begin{quote}
    \texttt{IInt\lbracket}\textit{kernels}\texttt{,}$x$\texttt{\rbracket}
\end{quote}
where $x$ might be a symbol or a numeric value and \textit{kernels} is a \texttt{List} of
\begin{quote}
    \texttt{omega\lbracket}$a$\texttt{\rbracket},\quad
    \texttt{tau},\quad
    \texttt{mu\lbracket}$n$\texttt{\rbracket},\quad
    \texttt{kappa},
\end{quote}
corresponding to the kernels in eq.~\eqref{eq:kernels} in the obvious way.

The package offers a few commands to the user:
\begin{itemize}
    \item \texttt{IIntInit\lbracket}\textit{prec}\texttt{\rbracket} \\
        Initialize \texttt{libiint} and set the working precision to \textit{prec} bits.
        Verbose output can be enabled with the option \texttt{Verbose->True}.
    \item \texttt{IIntCreate\lbracket}\textit{expression}\texttt{\rbracket} \\
        Register all \texttt{IInt}s in \textit{expression} and replace them by \texttt{IIntObj} objects.
    \item \texttt{IIntMatchEuclidean\lbracket}$q$\texttt{,}$a$\texttt{\rbracket} \\
        Match the constant terms of all registered \texttt{IInt}s along a path in the euclidean region from $1$ to $0$.
        For a point $x_1$ where the constant term is already known, the next point along the path $x_2$ is chosen so that the matching point $x$ fulfills $|x-x_1| \leq \min(qr_1,a)$ and $|x-x_2| \leq \min(qr_2,a)$, where $r_1$($r_2$) is the convergence radius for an expansion around $x_1$($x_2$).
    \item \texttt{IIntMatchPhysical\lbracket}$q$\texttt{,}$a$\texttt{\rbracket} \\
        Same as \texttt{IIntMatchEuclidean} but for a path in the physical region.
    \item \texttt{IIntSave\lbracket}\textit{filename}\texttt{\rbracket} \\
        Save all constant terms to \textit{filename} in the \texttt{YAML} file format.
    \item \texttt{IIntLoad\lbracket}\textit{filename}\texttt{\rbracket} \\
        Load all constant terms from \textit{filename} previously saved by \texttt{IIntSave}.
    \item \texttt{IIntEvaluate\lbracket}\textit{expression}\texttt{\rbracket} \\
        Evaluate all \texttt{IIntObj}s in \textit{expression} with numeric arguments \textit{x}.
        \textit{x} must lie within the convergence radius of the closest point where a constant term is available.
    \item \texttt{IIntSeries\lbracket}\textit{expression}\texttt{,\ltbrace}$x$\texttt{,}$x_0$\texttt{,}\textit{order}\texttt{\rtbrace\rbracket} \\
        Expand all \texttt{IIntOBj}s in \textit{expression} in a series in $(x-x_0)$ up to order \textit{order}.
\end{itemize}
In order to clarify the usage of the package, we refer to the files \texttt{match\_euclidean.m}, \texttt{match\_physical.m} and \texttt{eval.m} in the directory \texttt{eval/} of the \texttt{libiint} package.
These \texttt{Mathematica} scripts allow the evaluation of the coefficient $\tilde C_2^{(2)}$ of eq.~\eqref{eq:main-result} at various kinematic points.
The first two scripts are used to generate the \texttt{YAML} files containing the constant terms.
If one wants to evaluate the form factor only at the predefined points, this step that usually takes a few hours can be skipped since the package already includes the files \texttt{euclidean.yaml} and \texttt{physical.yaml} with all points irrelevant for the evaluation removed in order to save disk space.

The GPLs in eq.~\eqref{eq:main-result} are evaluated via \texttt{PolyLogTools}~\cite{Duhr:2019tlz} and \texttt{GiNaC}~\cite{Bauer:2000cp}, which have to be installed as well.
In order to run the scripts in the \texttt{eval/} directory, the file \texttt{ggh-nc2.m} (see section~\ref{appdx:suppl:result}) has to be copied (or symlinked) to the \texttt{eval/} directory.
The \texttt{eval.m} script is run with\footnote{The other two scripts are run likewise.}
\begin{quote}
    \texttt{math -script eval.m}
\end{quote}
and should produce an output of the form:

\vspace{.3cm}

\parbox{\textwidth}{
    \texttt{-0.00112 => 0.109734}[...]\texttt{~- 0.409664}[...]\texttt{*I (IInt: 0.46 sec., GPL: 13.4 sec.)} \\
    \texttt{0.0125 => -0.924519}[...]\texttt{~+ 0.406567}[...]\texttt{*I (IInt: 0.60 sec., GPL: 11.3 sec.)} \\
    \texttt{-0.75 => 0.3190702}[...]\texttt{~+ 5.7210056}[...]\texttt{*I (IInt: 0.66 sec., GPL: 341.2 sec.)} \\
}

The evaluation point $x$ is given at the far left, while the result evaluated at $x$ is given after the arrow.

\section{Analytic result} \label{appdx:result}
In this appendix, the analytic result for the coefficient $\tilde C^{(2)}_2$ is presented along with the two-loop contribution.
The three-loop result contains the three types of iterated integrals discussed in this article and defined in eqs.~(\ref{eq:gpl-def},\ref{eq:P},\ref{eq:iint},\ref{eq:kernels}).

In order to express the result in a more compact form, we introduce a few short-hand notations.
For the iterated integrals, we neglect the arguments, i.e.
\begin{align*}
    G_{a_1,\cdots,a_N} &\equiv G_{a_1,\cdots,a_N}(x)\,, \\
    P_{a_1,\cdots,a_N}^{n_1,\cdots,n_N} &\equiv P_{a_1,\cdots,a_N}^{n_1,\cdots,n_N}(\sqrt{x})\,, \\
    \xIInt{f_1\cdots f_N} &\equiv \IInt{f_1,\cdots,f_N}{1,x}\,.
\end{align*}
Furthermore, since the GPL integration kernels $1/(t-r)$ and $1/(t-r^*)$  as well as the kernels $1/(t-\varphi^2)$ and $1/(t-\varphi^{-2})$ always appear in pairs, we combine them with the help of two more integration kernels\footnote{Integration kernels of that form were proposed initially in Ref.~\cite{vonManteuffel:2013vja}.},
\begin{subequations}
    \begin{align}
        f_{\bar r}(t) &= \frac1{t-r} + \frac1{t-r^*} = \frac{2t - 1}{t^2-t+1}
        \,, \\
        f_{\bar \varphi}(t) &= \frac1{t-\varphi^2} + \frac1{t-\varphi^{-2}} = \frac{2t-3}{t^2-3t+1}
        \,,
    \end{align}
\end{subequations}
where $r$ denotes a sixth-root of unity $r=\mathrm{e}^{\mathrm{i}\pi/3}$ and $\varphi$ the golden ratio $\varphi = (1+\sqrt5)/2$; i.e. GPLs with some indices being $\bar r$ ($\bar\varphi$) can be interpreted as the sum of all possible GPLs, where $\bar r=r$ ($\bar\varphi = \varphi^2$) or $\bar r = r^*$ ($\bar\varphi = \varphi^{-2}$), e.g.
\begin{equation}
    G_{\bar r,1,\bar\varphi,0,\shortminus1,0}
    =
    G_{r,1,\varphi^2,0,\shortminus1,0}
    +
    G_{r,1,\varphi^{-2},0,\shortminus1,0}
    +
    G_{r^*,1,\varphi^2,0,\shortminus1,0}
    +
    G_{r^*,1,\varphi^{-2},0,\shortminus1,0}
    \,.
\end{equation}
The function $\psi(t)$, given in eq.~\eqref{eq:psi}, and its derivatives are abbreviated as
\begin{equation}
    \psi \equiv \psi(t)\,, \quad
    \psi' \equiv \frac{\d\psi(t)}{\d t}\,, \quad
    \psi'' \equiv \frac{\d^2\psi(t)}{\d t^2}\,.
\end{equation}
Moreover, Riemann's $\zeta$-function $\zeta_n = \sum_{j=1}^\infty j^{-n}$ and the golden ratio $\varphi=(1+\sqrt5)/2$ show up in the result and we define $L_\mu \equiv \ln(\mu^2/m_q^2)$, where $\mu$ denotes the renormalization scale.

The two-loop result, which has been known already for 15 years~\cite{Harlander:2005rq,Aglietti:2006tp,Anastasiou:2006hc}, reads in our notation,
\begin{subequations}
    {\allowdisplaybreaks\begin{align}
        \begin{split} \label{eq:result:C1-1}
            \tilde C_1^{(\shortminus1)} &=
            -\frac{x(5x^{2}-6x+5)}{2(x-1)^{4}}G_{1,0,0}+\frac{x(x+1)(x^{2}+1)}{(x-1)^{5}}\bigg(-4G_{0,0,\shortminus1,0}-2\zeta_3G_{0}
            \\&\quad
            -\frac{1}{40}\pi^{4}-\frac{1}{6}\pi^{2}G_{0,0}-\frac{1}{4}G_{0,0,0,0}-G_{0,0,1,0}+2G_{0,\shortminus1,0,0}+\frac{7}{2}G_{0,1,0,0}\bigg)
            \\&\quad
            +\frac{x(x+1)^{2}}{(x-1)^{4}}\bigg(\frac{1}{2}G_{0,1,0}+\frac{1}{12}\pi^{2}G_{0}+2G_{0,\shortminus1,0}\bigg)+\frac{5x}{2(x-1)^{2}}-\frac{3x(x+1)}{2(x-1)^{3}}G_{0}
            \\&\quad
            -\frac{x(x^{2}-14x+1)}{2(x-1)^{4}}\zeta_3+\frac{x(3x^{3}-7x^{2}+25x+3)}{4(x-1)^{5}}G_{0,0,0}-\frac{3x^{2}}{(x-1)^{4}}G_{0,0}
            \,,
        \end{split}
        \\
        \begin{split} \label{eq:result:C10}
            \tilde C_1^{(0)} &= n_l\bigg[\frac{x}{3(x-1)^{2}}\bigg(-2G_{1}+L_\mu+G_{0}\bigg)+\frac{x(x+1)^{2}}{(x-1)^{4}}\bigg(-\frac{1}{2}G_{0,0,0}+\frac{1}{3}G_{0,0,1}
            \\ &\quad\qquad
            -\frac{1}{6}G_{0,0}L_\mu+\frac{1}{3}G_{0,1,0}+\frac{1}{3}G_{1,0,0}\bigg)\bigg]
            \,,
        \end{split}
        \\
        \begin{split} \label{eq:result:C11}
            \tilde C_1^{(1)}&=\frac{11x}{(x-1)^{2}}\bigg(-\frac{1}{2}-\frac{1}{6}L_\mu+\frac{1}{3}G_{1}\bigg)+\frac{x(x+1)^{2}}{(x-1)^{4}}\bigg(-\frac{11}{6}G_{0,0,1}-\frac{7}{3}G_{0,1,0}
            \\ &\quad
            -6\zeta_3 G_{1}-4G_{1,0,\shortminus1,0}-2G_{0,\shortminus1,0}-\frac{1}{3}\pi^{2}G_{1,0}-\frac{1}{12}\pi^{2}G_{0}+2G_{1,0,0,0}
            \\ &\quad
            +\frac{11}{12}G_{0,0}L_\mu\bigg)+\frac{x(x+1)(x^{2}+1)}{(x-1)^{5}}\bigg(-\frac{7}{2}G_{0,1,0,0}-2G_{0,\shortminus1,0,0}+G_{0,0,1,0}\bigg)
            \\ &\quad
            +\frac{x(x+1)(25x^{2}-7)}{360(x-1)^{5}}\pi^{4}-\frac{x(2x^{3}+7x^{2}+20x+7)}{2(x-1)^{5}}G_{0,0,0}-\frac{x(x-10)}{3(x-1)^{3}}G_{0}
            \\ &\quad
            +\frac{x(x+1)(3x^{2}-1)}{4(x-1)^{5}}G_{0,0,0,0}+\frac{x(x+1)(5x^{2}-1)}{(x-1)^{5}}\zeta_3G_{0}+\frac{x^{3}(x+1)}{3(x-1)^{5}}\pi^{2}G_{0,0}
            \\ &\quad
            +\frac{x(x^{2}+8x+1)}{2(x-1)^{4}}G_{0,0}+\frac{2x(x+1)(3x^{2}+1)}{(x-1)^{5}}G_{0,0,\shortminus1,0}+\frac{x(7x^{2}+6x+7)}{2(x-1)^{4}}\zeta_3
            \\ &\quad
            +\frac{x(11x^{2}+10x+11)}{3(x-1)^{4}}G_{1,0,0}
            \,,
        \end{split}
    \end{align}}%
\end{subequations}%
and finally, the main result of this article, i.e. the leading color contribution at three-loop, reads,
{\allowdisplaybreaks\tiny
}

\bibliography{biblio}

\end{document}